\documentclass[12pt]{article}
\usepackage[dvips]{graphicx}
\usepackage{amsmath}
\usepackage{slashed}

\newcommand{\aeq}{\begin{equation}}
\newcommand{\ceq}{\end{equation}}
\newcommand{\aec}{\begin{eqnarray}}
\newcommand{\cec}{\end{eqnarray}}
\newcommand{\ase}{\begin{subequations}}
\newcommand{\cse}{\end{subequations}}

\renewcommand{\(}{\left(}
\renewcommand{\)}{\right)}

\renewcommand{\a}{\alpha}
\renewcommand{\b}{\beta}

\newcommand{\m}{\mu}
\newcommand{\n}{\nu}
\renewcommand{\o}{\omega}
\newcommand{\g}{\gamma}
\renewcommand{\d}{\delta}
\newcommand{\h}{\eta}

\newcommand{\y}{\psi}
\renewcommand{\l}{\lambda}
\newcommand{\s}{\sigma}
\renewcommand{\r}{\rho}
\renewcommand{\k}{\kappa}
\newcommand{\q}{\theta}
\newcommand{\e}{\epsilon}
\newcommand{\x}{\xi}

\newcommand{\p}{\pi}

\newcommand{\G}{\Gamma}
\renewcommand{\P}{\Pi}

\newcommand{\D}{\Delta}

\newcommand{\pd}{\partial}

\begin{document}
\begin{center}
{\Large \bf  Bosonic and fermionic Weinberg-Joos $(j,0)\oplus (0,j)$ states of arbitrary spins 
as Lorentz-tensors or tensor-spinors and second order theory   }
\end{center}
\vspace{0.53cm}
\centerline{\large E. G. Delgado Acosta, V.M. Banda Guzm\'an, and M. Kirchbach }
\vspace{0.23cm}

\begin{center}
{Instituto de Fisica, UASLP,
Av. Manuel Nava 6, Zona Universitaria,\\
San Luis Potosi, SLP 78290, M\'exico
}
\end{center}

\vspace{0.53cm}
\begin{abstract}{
We propose a general method for the description of arbitrary single spin-$j$ states transforming according to $(j,0)\oplus(0,j)$ carrier spaces of the Lorentz algebra in terms  of Lorentz-tensors for bosons, and tensor-spinors for
 fermions, and by means of second order  Lagrangians. The method allows to avoid the cumbersome  matrix calculus and  higher
$\partial^{2j}$  order wave equations inherent to the Weinberg-Joos approach.
We start with reducible  Lorentz-tensor (tensor-spinor) representation spaces hosting  one sole
$(j,0)\oplus(0,j)$ irreducible sector and design there a representation reduction algorithm based on one of  the
Casimir invariants of the Lorentz algebra. This algorithm  allows us to
separate neatly the pure spin-$j$ sector of interest from the rest, while preserving the separate Lorentz-- and Dirac indexes.
However, the Lorentz invariants are momentum independent and do not provide wave equations.
Genuine wave equations are obtained by conditioning   the Lorentz-tensors under consideration to satisfy the Klein-Gordon equation. 
In so doing, one always ends up with wave equations and associated Lagrangians that are second order in the momenta.
Specifically, a spin-3/2 particle transforming as $(3/2,0)\oplus (0,3/2)$ is comfortably  described by a  second order Lagrangian in the basis
of the totally  antisymmetric Lorentz tensor-spinor of
second rank, $\Psi_{\left[ \mu\nu\right]}$. Moreover, the particle is shown to propagate causally within an electromagnetic background. 
In our study of $(3/2,0)\oplus(0,3/2)$ as part of $\Psi_{\lbrack \mu\nu\rbrack }$ we  reproduce the
electromagnetic multipole moments known from the Weinberg-Joos theory. We also
find  a Compton differential cross section that satisfies unitarity in forward direction.
The suggested tensor calculus presents itself very computer friendly with respect to the symbolic software 
FeynCalc.}
\end{abstract}

\begin{flushleft}
{{PACS-key}: {11.30.Cp (Lorentz and Poincar\'e invariance),  03.65.Pm
(Relativistic wave equations)} 
     } 
\end{flushleft}

\def\s{\mbox{\boldmath$\displaystyle\mathbf{\sigma}$}}
\def\J{\mbox{\boldmath$\displaystyle\mathbf{J}$}}
\def\K{\mbox{\boldmath$\displaystyle\mathbf{K}$}}
\def\P{\mbox{\boldmath$\displaystyle\mathbf{P}$}}
\def\p{\mbox{\boldmath$\displaystyle\mathbf{p}$}}
\def\hp{\mbox{\boldmath$\displaystyle\mathbf{\widehat{\p}}$}}
\def\x{\mbox{\boldmath$\displaystyle\mathbf{x}$}}
\def\0{\mbox{\boldmath$\displaystyle\mathbf{0}$}}
\def\bv{\mbox{\boldmath$\displaystyle\mathbf{\varphi}$}}
\def\hbv{\mbox{\boldmath$\displaystyle\mathbf{\widehat\varphi}$}}

\def\bg{\mbox{\boldmath$\displaystyle\mathbf{\gamma }$}}

\def\bl{\mbox{\boldmath$\displaystyle\mathbf{\lambda}$}}
\def\br{\mbox{\boldmath$\displaystyle\mathbf{\rho}$}}
\def\1{\mbox{\boldmath$\displaystyle\mathbf{1}$}}
\def\bfhh{\mbox{\boldmath$\displaystyle\mathbf{(1/2,0)\oplus(0,1/2)}\,\,$}}

\def\mn{\mbox{\boldmath$\displaystyle\mathbf{\nu}$}}
\def\amn{\mbox{\boldmath$\displaystyle\mathbf{\overline{\nu}}$}}

\def\mne{\mbox{\boldmath$\displaystyle\mathbf{\nu_e}$}}
\def\amne{\mbox{\boldmath$\displaystyle\mathbf{\overline{\nu}_e}$}}
\def\rlh{\mbox{\boldmath$\displaystyle\mathbf{\rightleftharpoons}$}}

\def\wm{\mbox{\boldmath$\displaystyle\mathbf{W^-}$}}
\def\hh{\mbox{\boldmath$\displaystyle\mathbf{(1/2,1/2)}$}}
\def\h00h{\mbox{\boldmath$\displaystyle\mathbf{(1/2,0)\oplus(0,1/2)}$}}
\def\znbb{\mbox{\boldmath$\displaystyle\mathbf{0\nu \beta\beta}$}}


\vspace{1truecm}

\section{Introduction}\label{sec1}


Particles of high-spins  \cite{Weinberg:1995mt} continue being among the most
enigmatic  challenges in contemporary
theoretical physics. The difficulties in their descriptions, both at the
classical-, and the quantum-field theoretical
levels, are well known and take their origin from the circumstance  that such
particles are most naturally described
by  differential equations of orders  twice their  respective  spins
\cite{Weinberg64}-\cite{Eeg3}. Higher-order theories are  difficult to tackle
and various strategies have been elaborated over the years to lower the order
of the corresponding differential equations, the linear
ones by Rarita-Schwinger \cite{RS} being the most popular so far. However, the
latter framework is plagued by various
inconsistencies, the acausal propagation within an electromagnetic
environment \cite{Velo:1970ur}, the violation of unitarity in
Compton scattering in the ultraviolet in schemes with minimal gauge couplings
\cite{Ferrara},
and the violation of Lorentz-symmetry  upon quantization, being the most
serious ones.
In parallel, also second order spin-$\frac{1}{2}$ \cite{Hostler},
\cite{Morgan}, \cite{Vaquero}, \cite{Rene} and spin-$\frac{3}{2}$
\cite{Napsuciale:2006wr}
fermion theories have been developed  by different authors and shown to
provide a reasonable compromise between the rigorous linear--
and the  natural higher-order descriptions in so far as they were able to
circumvent, among others,  the acausality problem
and the related violation of unitarity in Compton scattering
\cite{DelgadoAcosta:2009ic}. However,
for spins higher than $\frac{3}{2}$ no second order theory has been developed
so far. It is the goal of the present work to fill this gap.
The interest in such a study is motivated by the observation that
distinct representation spaces  of the Lorentz algebra $so(1,3)$ describe
particles of different physical properties.
{}For example, due to the representation dependence of the boost operator,
the  electromagnetic quadrupole and octupole moments of fundamental particles
with spin-$\frac{3}{2}$ transforming
in the four-vector spinor come out  different from those of particles 
transforming  as
$\left( \frac{3}{2},0\right) \oplus \left(0,\frac{3}{2} \right)$
\cite{DelgadoAcosta:2012yc}.
Same holds valid regarding spin-$1$ in the four-vector, $\left( \frac{1}
{2},\frac{1}{2}\right)$,
versus the anti-symmetric tensor, $\left( 1,0\right)\oplus\left( 0,1\right)$ \cite{DelgadoAcosta:2013}.
In view of the expected production of new particles in the experiments run by
the Large Hadron Collider
it is important to have at ones disposal a reliable and comfortable to deal
with universal calculation scheme
for high spins transforming in carrier spaces of the Lorentz algebra different from the totally symmetric tensors 
of common use, the Weinberg-Joos states being the prime candidates.  The present study is devoted
to the elaboration of such a scheme.\\

\noindent
The  path we take is to  embed single  spin-$j$ Weinberg-Joos
states \cite{Weinberg64}-\cite{Eeg3}, $(j,0)\oplus (0,j)$,
into direct sums of properly selected irreducible $so(1,3)$ representation
spa\-ces \-which join to reducible representation spaces that are large enough as to
allow to be equipped by Lorentz, and if needed, separate Dirac indexes.
Then we pin down  the state of our interest in a two step procedure.
First  we pin down the $(j,0)\oplus(0,j)$ irreducible representation space by means of a
momentum independent (static)
projector designed on the basis of one of the Casimir invariants of the Lorentz
algebra and then impose on it the Klein-Gordon equation. 
In this fashion, a second order formalism for any single-spin valued Weinberg-
Joos state is furnished. The scheme also allows for an extension to
spin-$j$ as the highest spins in the two-spin valued representation spaces, 
$ \left(j-\frac{1}{2},\frac{1}{2}\right)\oplus \left(\frac{1}{2}, j-\frac{1}{2}\right)$.\\

\noindent
The paper is organized as follows. In the next section we formulate  the concepts underlying  our
suggested method. In section III we design the calculation algorithm and write down a second order
master equation for any single spin-$j$ transforming as $(j,0)\oplus (0,j)$.
In same section we apply the scheme elaborated to the description of the  particular case of spin-$\frac{3}{2}$.
In section IV we present  the  spin-$\frac{3}{2}$
Lagrangian and couple it to the electromagnetic field, find the electromagnetic current, and
calculate the associated electromagnetic multipole moments. We show that the
observables obtained in this fashion
reproduce  the predictions reported in \cite{DelgadoAcosta:2012yc} where the
pure  spin-$\frac{3}{2}$ states have been considered in the standard way as eight-dimensional spinors. Also there we show that the pure spin-$\frac{3}{2}$ sector of the
antisymmetric tensor spinor describes a particle that propagates causally within an electromagnetic environment.
Finally, we calculate the process of Compton scattering off  $\left(
\frac{3}{2},0\right)\oplus \left(0,\frac{3}{2} \right)$ and report on a finite differential cross section in forward direction. 
The paper closes with brief conclusions and has one Appendix.



\section{ Lorentz tensors  and  tensor-spinors for bosons and fermions of  any spin and second order wave equations }

The method for high-spin description advocated in this work is based upon
representation spaces of the Lorentz algebra $so(1,3)$, which are as a rule
different from those of common use.
While the representation spaces underlying the Weinberg-Joos formalism are of
non-tensorial nature,
those underlying the Rarita-Schwinger framework are totally symmetric tensors.
Our method permits to consider besides totally symmetric also 
totally antisymmetric Lorentz tensors,  and also such of mixed symmetries.
Our idea is to embed  $(j,0)\oplus (0,j)$ carrier spaces of $so(1,3)$
into finite direct sums of  properly chosen dummy irreducible
representation spaces with the
aim to end up with reducible representation spaces of  minimal dimensionality  which are nonetheless
large enough as to allow  to be equipped by  Lorentz-- (and if needed,
separate  Dirac) indexes, i.e.
\begin{eqnarray}
   \Psi_{\mu_1,..\mu_t}&\simeq&
\left[(j,0)\oplus (0,j)\right] \oplus
\Sigma _{(k,l)} \, n_{(kl)} \left[
\left(j_k,j_l\right)\oplus \left(j_l,j_k\right)\right].\nonumber\\
\label{chudo_1}
\end{eqnarray}

{}For example, pure spin-$\frac{3}{2}$ can be embedded into the totally
antisymmetric
tensor of second rank, $B_{\left[\mu \nu \right]}$,  with Dirac spinor components,
$\psi$, a representation space that is reducible
according to,
\begin{eqnarray}
\Psi_{\left[ \mu\nu\right] }&\simeq&B_{\left[\mu\nu \right]}\otimes \psi\nonumber\\
&\simeq&
\left[ (1,0)\oplus (0,1)\right]\otimes \left[\left(\frac{1}{2},0\right)\oplus
\left(0, \frac{1}{2} \right) \right]\nonumber\\
&\longrightarrow &
   \left[\left(
\mathbf{ \frac{3}{2} },{\mathbf 0}\right)\oplus \left({\mathbf 0},
\mathbf{\frac{3}{2}} \right)\right]\oplus 
\left[\left(\frac{1}{2},0\right)
\oplus \left(0, \frac{1}{2} \right)\right]\nonumber\\
&\oplus&
\left[\left( 1,\frac{1}{2}\right)\oplus \left(\frac{1}{2}, 1 \right) \right]
\nonumber\\
&\longrightarrow & 
\left[\left( \mathbf{ \frac{3}{2} },{\mathbf 0}\right)\oplus \left({\mathbf 0},
\mathbf{\frac{3}{2}} \right)\right] \oplus \psi_\mu,\nonumber\\
\psi_\mu &\simeq&  \left(\frac{1}{2},\frac{1}{2}\right) \otimes\left[ \left(\frac{1}{2},0 \right) \oplus \left(0,\frac{1}{2}
\right)\right],
\label{tensor_spinor}
\end{eqnarray}
where we use $\left[\mu\nu\right]$ to denote an antisymmetric pair of indexes. 
Spin-$2$ is part of the antisymmetric tensor with anti-symmetric tensor components, 
\begin{eqnarray}
\Phi_{\left[ \mu\nu\right]\left[\eta \gamma \right] }&\simeq&B_{\left[\mu\nu \right]}\otimes B_{\left[\eta\gamma \right]}\nonumber\\
&=&\left[ (1,0)\oplus (0,1)\right]\otimes \left[\left(1,0\right)\oplus \left(0,1\right)\right] \nonumber\\
&\longrightarrow &
\left[\left({\mathbf 2},0 \right)\oplus \left(0,{\mathbf 2} \right)\right]\oplus 2\left(0,0 \right)\oplus
2\left(1,1 \right)\nonumber\\
&\oplus& \left[ \left(1,0 \right)\oplus \left(0,1 \right)\right],
\label{tensor_vector}
\end{eqnarray}
where the numbers in front of the irreps indicate their multiplicity upon
reduction.\\
Similarly, spin-$\frac{5}{2}$ can be viewed as a resident of the direct product of the
antisymmetric tensor-tensor
from (\ref{tensor_vector}) with a Dirac spinor, giving rise to $\Psi_{\left[\mu\nu \right]\left[\eta\gamma\right]}$, a
representation space reducible according to
\begin{eqnarray}
\Phi_{\left[\mu\nu \right]\left[ \eta\gamma \right]} &\otimes& \psi \simeq
\Psi_{\left[ \mu \nu\right]\left[ \eta \gamma\right] }\nonumber\\
&\simeq&
\left[ (1,0)\oplus (0,1)\right]\otimes
\left[ (1,0)\oplus(0,1) \right]\nonumber\\
&\otimes& 
\left[\left(\frac{1}{2},0\right)\oplus \left(0, \frac{1}{2} \right) \right]
\nonumber\\
&\longrightarrow &
\left[\left(\mathbf{\frac{5}{2}},0\right)\oplus \left(0, \mathbf{\frac{5}{2}} \right)
\right]\oplus
3\left[\left( 1,\frac{1}{2}\right)\oplus \left(\frac{1}{2}, 1 \right)
\right]\nonumber\\
&\oplus& 2\left[\left(1,\frac{3}{2}\right)\oplus \left(\frac{3}{2}, 1 \right)
\right]
\oplus 3\left[\left( \frac{1}{2},0\right)\oplus \left(0,\frac{1}{2} \right)
\right]\nonumber\\
&\oplus& 2\left[\left( \frac{3}{2},0\right)\oplus \left(0,\frac{3}{2} \right) \right]
\oplus \left[\left(2,\frac{1}{2}\right)\oplus \left(\frac{1}{2}, 2\right) \right].\nonumber\\
\label{tensor_4Vspinor}
\end{eqnarray}

\noindent
In order to pin down the $(j,0)\oplus((0,j)$ sector of our interest we have to  remove the  dummy irreducible sectors
in the above reducible tensorial representation spaces.  For this purpose we develop an algorithm on the basis of
static projectors constructed from one of the Casimir invariants of the Lorentz
algebra. Such projectors have  the property to unambiguously identify and exclude
anyone of the irreducible representation spaces, no matter whether single- or
multiple-spin valued  \cite{Wyborne}.
Below we outline this representation reduction algorithm.

\subsection{A Casimir invariant of the Lorentz algebra and the  $(j,0)\oplus(0,j)$ tracking algorithm }
The Lorentz algebra has a Casimir operator,  denoted by $F$ and given in \cite{Wyborne} in terms of the Lorentz-group generators, 
$M_{\mu\nu}$, as
\begin{eqnarray}
~[F]_{AB}&=&\frac{1}{4}[M^{\m\n}]_A{}^C
[M_{\m\n}]_{C B},
\label{FG_C}
\end{eqnarray}
with the capital Latin letters  $A$,$B$, $C$, ... standing for the generic indexes characterizing the
representation space of interest. Its eigenvalue problem for any irreducible representation
spaces of the type $ (j_1,j_2)\oplus (j_2,j_1)$, with the generic representation functions denoted by, $w^{(j_1,j_2)}$,
reads,
\begin{eqnarray}
F \,w^{(j_1,j_2)}&=&c_{(j_1,j_2)}w^{(j_1,j_2) },\nonumber\\
c_{(j_1,j_2)}&=&j_1(j_1+1)+j_2(j_2+1)
=\frac{1}{2}\left(K(K+2)+M^2 \right)
\label{FCasm}
\end{eqnarray}
 where
\begin{eqnarray}
K=j_1+j_2, \quad M=|j_1-j_2|.
\label{Fev}
\end{eqnarray}

On the basis of  $F$ we design the  following momentum independent Lorentz
projector, $\mathcal{P}_F^{(j_1,j_2)}$,
\begin{eqnarray}
\mathcal{P}^{(j_1,j_2)}_F w^{(j_1,j_2)} &=&\left[\Pi_{kl}\times  \left(
\frac{
F-c_{(j_k,j_l)}
}{c_{(j_1,j_2)}-c_{(j_k,j_l)}}\right)\right]w^{(j_k,j_l)}
=w^{(j_1,j_2)},
\label{feq12}
\end{eqnarray}
where $ \Pi_{kl}\times$ denotes the operation of successive multiplication, $c_{(j_1,j_2)}$ is the $F$ eigenvalue of the searched sector, 
$(j_1,j_2)\oplus (j_2,j_1)$, while
$c_{(j_k,j_l)}$ are the $F$ eigenvalues of the dummy sectors, $(j_k,j_l)\oplus (j_l,j_k)$,  of the hosting tensor,  which need all to be 
 excluded.
The mayor advantage of such projectors is that they are  of zeroth order in the momenta,
and  do not contribute at all to  the order of the wave equations.
In what follows we shall mainly consider only such reducible Lorentz tensors
(or, tensors-spinors)
which allow the spin-$j$ of our interest to reside within one single-spin valued
irreducible subspace, $(j,0)\oplus(0,j)$,
i.e. in 
\begin{equation}
(j_1,j_2)\oplus(j_2,j_1):\quad  j=j_1,\quad  j_2=0.
\label{j00j}
\end{equation}
However, in the next section we show that the algorithm suggested allows for an extension also toward 
double-spin valued spaces such as  $j, j^\prime \in (j_1,j_2)\oplus (j_2, j_1)$ with $j_2=\frac{1}{2}$, i.e.  
\begin{equation}
\left(j_1, \frac{1}{2} \right)\oplus \left(\frac{1}{2},j_1 \right):\quad {\Big\{}
\begin{array}{cc}
   j=j_1-\frac{1}{2},&\\
j^\prime=j_1+\frac{1}{2}.&
\end{array}
\label{chudo_2}
\end{equation}

\subsection{Second order master equations for any pure spin }
The dynamics into the spin-$j$ sectors of interest from above is introduced by implementing the mass-shell condition,
\begin{eqnarray}
\frac{P^2}{m^2} \, w^{(j,0)}
&=&w^{(j,0)},\label{p2j0}
\end{eqnarray}
for the case of single spins transforming as $(j,0)\oplus(0,j)$.
{}For double-spin valued representation spaces, the Lorentz projector only tracks down the 
$\left(j_1,\frac{1}{2}\right)\oplus \left(\frac{1}{2},j_1\right)$ sector as a whole, but does not distinguish between its  
$j=\left(j_1-\frac{1}{2}\right)$, and $j^\prime =\left(j_1+\frac{1}{2}\right)$
residents. In order to single out, say, the spin-$j$, a different projector, here denoted by 
$\mathcal{P}^{(m,j)}_{{\mathcal W}^2}(p)$, and based on the squared Pauli-Lubanski vector, has to be employed.
The mass projector, $P^2/m^2$, and the spin-$j$ projector, $\mathcal{P}^{(m,j)}_{{\mathcal W}^2}(p)$, will be occasionally referred  to as
``Poincar\'e projectors'' ~\cite{Napsuciale:2006wr} in reference to the fact that they  express in terms of
the  Casimir invariants of the Poincar\'e algebra, the squared four momentum,
$P^2$, and the squared Pauli-Lubanski vector, ${\mathcal W}^2(p)$, as
\begin{eqnarray}
\mathcal{P}^{(m,j)}_{{\mathcal W}^2}(p) \,   w^{\left(j_1,\frac{1}{2}\right);j}&=&
w^{\left(j_1,\frac{1}{2}\right);j},\label{w2eq120}\\
\mathcal{P}^{(m,j)}_{{\mathcal W}^2}(p)&=&\frac{P^2}{m^2} \frac{{\mathcal W}^2(p)-\e_{j^\prime}}{\e_j-\e_{j^\prime}}.
\end{eqnarray}
Here, ${\mathcal W}^\mu(p)$ denotes the Pauli-Lubanski (pseudo)vector,
defined as
\begin{equation}
\left( {\mathcal W}^{\mu}\right)_{AB}(p)=\frac{1}{2}\epsilon_{\lambda
\rho\sigma\mu}\left(M^{\rho\sigma} \right)_{AB}p^\mu,
\label{gen_PL}
\end{equation}
where $M^{\rho\sigma}$ are the generators of the Lorentz algebra in the
representation space of interest,
while $A$, and $B$ are again as already mentioned  above the sets of indexes that  characterize the
dimensionality of that very representation space. Furthermore,
$\e_{j}=-p^2 j(j+1)$, and $\epsilon_{j^\prime}=-p^2 j^\prime (j^\prime +1)$, are  the respective eigenvalues of the
eigenstates of the operator ${\mathcal W}^2(p)$,  corresponding to the spins-$j$, and $j^\prime$,  residing within 
$\left(j_1,\frac{1}{2}\right)\oplus \left(\frac{1}{2},j_1\right)$, with the common mass-$m$ satisfying $m^2=p^2$.\\

\noindent
Combining now the equation (\ref{feq12}) either with  (\ref{p2j0}), or with (\ref{w2eq120})--(\ref{gen_PL}),
the following master equations  emerge,
\begin{eqnarray}
\mbox{for}\quad j\in (j,0)\oplus (0,j):&&
\left[\frac{P^2}{m^2} \mathcal{P}^{(j,0)}_F \right]_{B}{}^{A}\,  w^{(j,0)}_{A}=w^{(j,0)}_{B},
\label{masterequations_1}
\end{eqnarray}
and

\begin{eqnarray}
\mbox{for} \, \, j\in \left(j_1,\frac{1}{2}\right)\oplus \left( \frac{1}{2},j_1\right): &&
\left[\frac{P^2}{m^2} \mathcal{P}^{(m,j)}_{{\mathcal W}^2}(p)
\mathcal{P}^{\left(j_1,\frac{1}{2}\right)}_F \right]_{D}{}{}^{C}\,
w^{\left(j_1,\frac{1}{2}\right);j}_{C} 
=w^{\left(j_1,\frac{1}{2}\right);j}_{D},\nonumber\\
\label{masterequations_2}
\end{eqnarray}
respectively.
Here, the functions, $w^{(j,0)}$, and 
$w^{\left(j_1,\frac{1}{2}\right);j}$,  have the property each  to simultaneously diagonalize both the Lorentz and the Poincar\'e projectors.
The indexes $A$ for bosons are given by, $A={\mu_1,\mu_2...\mu_t}$, while those for fer\-mi\-ons carry in addition a Dirac index, 
denoted by small Latin letters, $a$, $b$, etc. according to $A={\mu_1\mu_2...\mu_t};a$.
Along this path,  one necessarily  encounters Lagrangians that are second
order in  the momenta.

\noindent
Second order fermion approaches  have  traditions in field  theory
\cite{Hostler},\cite{Morgan},
and are of growing popularity in QED as well as in QCD \cite{Schubert},
\cite{Krasnov}, \cite{DelgadoAcosta:2010nx}.

\noindent
The present work focuses on the description of the pure spin-$\frac{3}{2}$
Weinberg-Joos state,
$\left(\frac{3}{2}, 0 \right)\oplus \left(0,\frac{3}{2} \right)$ as part of
the antisymmetric tensor-spinor of second rank
in (\ref{tensor_spinor}).\\

\noindent
{}Fermionic representation spaces of the types (\ref{tensor_spinor}), (\ref{tensor_4Vspinor})  
have been earlier employed  by Niederle and Nikitin in \cite{Niederle} in a linear framework of the  Rarita-Schwinger type,  
with the special  emphasis on spin-$\frac{3}{2}$.
However, the focus of \cite{Niederle} has been in first place the separation of parities, while 
the question on the precise  assignment of spin-$\frac{3}{2}$ to the
irreducible $\left( \frac{3}{2},0\right) \oplus \left(0,\frac{3}{2} \right)$
sector of the anti-symmetric tensor spinor of second rank has been  left aside. 
Moreover, differently from the present  work, no physical observables have been calculated in \cite{Niederle} 
for the sake of a comparison to  predictions by other formalisms.
{}Finally, the scheme of \cite{Niederle} confines to fractional spins alone, while the one elaborated here 
applies to both fermions and bosons.

\section{Pure spin-$\frac{3}{2}$ in $\left(\frac{3}{2},0\right)\oplus\left(0,\frac{3}{2}\right)$ as part of the anti-symmetric Lorentz tensor-spinor of second rank}

\subsection{The anti-symmetric Lorentz  tensor spinor of second rank}
The antisymmetric Lorentz tensor-spinor of second rank has been defined in eq.~(\ref{tensor_spinor}) and  
is the ordinary anti-sym\-met\-ric Lorentz-tensor of second rank , $(1,0)\oplus(0,1)$, with Dirac spinor, $\left(\frac{1}{2},0\right)\oplus \left(0,\frac{1}{2}\right)$, 
components. The $(1,0)\oplus (0,1)$ sector is well known and has been frequently elaborated in the literature, listed among  others in
\cite{Dadich2002},\cite{DelgadoAcosta:2012yc}. We here present it in the momentum space, and denote it by,
${\mathcal B}^{\left[ \alpha \beta \right]}({\mathbf p},\lambda^\prime )$ with ${\mathbf p}$ standing for the three momentum, 
and $\lambda^\prime $ denoting the polarization quantum number, taking the values $\lambda^\prime =\pm 1,0$.
The spin-$1$ tensor ${\mathcal B}^{\left[ \alpha \beta \right]}({\mathbf p},\lambda^\prime )$ allows for a representation in terms of the three
spin-$1$ basis states, $\eta^\alpha({\mathbf p}, 1, \lambda^\prime )$, and the one spin-$0$ state, 
$\eta ^\alpha ({\mathbf p},0,0)$, spanning the four-vector space, $\left(\frac{1}{2},\frac{1}{2}\right)$. 
These states have been constructed for example in \cite{Ahluwalia:2001}, \cite{DelgadoAcosta:2012yc} and are summarized   
in the equations (\ref{eta1pl})--(\ref{eta0}) in the Appendix below. In terms of the aforementioned  $\left(\frac{1}{2},\frac{1}{2}\right)$ basis vectors, 
the tensor under discussion expresses as,
\begin{equation}
{\mathcal B}^{\left[ \alpha\beta \right]}({\mathbf p},\lambda^\prime )=\eta ^\alpha ({\mathbf p}, 0,0)\eta^\beta ({\mathbf p},1,\lambda^\prime ) -
\eta^\alpha ({\mathbf p},1,\lambda^\prime )\eta ^\beta ({\mathbf p}, 0,0). 
\end{equation}  

Now the tensor-spinor of interest here, to be denoted by 
${\mathcal T}_\pm ^{\left[\alpha\beta \right]}({\mathbf p},\lambda^\prime , \lambda^\prime{}^\prime )$, its dual being
${\widetilde {\mathcal T}}_\pm ^{\left[\alpha\beta \right]}({\mathbf p},\lambda^\prime , \lambda^\prime{}^\prime )$, reads,
\begin{eqnarray}
{\mathcal T}^{\left[\alpha\beta \right]}_\pm ({\mathbf p},\lambda^\prime , \lambda^\prime{}^\prime  )&=&
{\mathcal B}^{\left[ \alpha\beta\right]}({\mathbf  p},\lambda^\prime )\otimes u_\pm ({\mathbf p},\lambda^\prime{}^\prime ),
\label{TS_MSP}
\end{eqnarray}
where $u_+({\mathbf p},\lambda^\prime{}^\prime )$ and  $u_-({\mathbf p},\lambda^\prime{}^\prime )$ denote in their turn the 
$u ({\mathbf p},\lambda^\prime{}^\prime )$ and the $v ({\mathbf p},\lambda^\prime{}^\prime )$ Dirac spinors of positive/negative parities,
and $\lambda^\prime{}^\prime =\pm \frac{1}{2}$.
This tensor is reducible according to (\ref{tensor_spinor}) and its irreducible  sectors can be identified by means of
projectors constructed from one of the Casimir invariants of the Lorentz algebra, 
an issue on which we shed light in subsection 3.3 below. Before that, in the subsequent section we present the generators of
the Lorentz algebra in the anti-sym\-met\-ric ten\-sor-spi\-nor of se\-cond rank.  

\subsection{The Lorentz algebra generators in the anti-symmetric tensor spinor}
The generators within the anti-symmetric tensor-spinor (ATS) of second rank are 
\begin{eqnarray}
[M^{ATS}_{\m\n}]_{\left[ \a\b\right]\left[\g\d\right]}&=&[M^{AT}_{\m\n}]_{\left[\a\b\right]\left[\g\d\right]}\mathbf{1}^S+\mathbf{1}_{\left[\a\b\right]\left[\g\d\right]}\,\,
\left[M^S_{\m\n}\right],
\label{def:gensATS} \\
\mathbf{1}_{\left[\a\b\right]\left[\g\d\right]}&=&\frac{1}{2}(g_{\a\g}g_{\b\d}-g_{\a\d}g_{\b\g}),\\ 
M^{S}_{\m\n}&=&\frac{1}{2}\s_{\m\n}=\frac{i}{4}[\g_\m,\g_{\n}].\label{genss}
\label{identityAT}
\end{eqnarray}
Here, $[M^{AT}_{\m\n}]_{\left[\a\b\right]\left[\g\d\right]}$ are the generators within the anti-symmetric tensor (AT) space,
$\mathbf{1}_{\left[\a\b\right]\left[\g\d\right]}$ stands for the identity in this space, while $\mathbf{1}^S$ and 
$M^S_{\m\n}$ are the unit operator and the generators within the Dirac space (\ref{genss}), where  $\g_\m$ are the standard  Dirac  matrices.
In what follows we shall suppress the Dirac indexes for the sake of avoiding cumbersome notations 
and will keep only the Lorentz indexes. We always will mark anti-symmetric pairs of Lorentz indexes by 
$\left[...\right]$ any times when  more than one pair is involved.

The $[M^{AT}_{\m\n}]_{\left[\a\b\right]\left[\g\d\right]}$ generators express in terms of the generators in the four-vector, 
\begin{eqnarray}
~[M^{V}_{\m\n}]_{\a\b}&=&i(g_{\a\m}g_{\b\n}-g_{\a\n}g_{\b\m}),\label{gensv}\
\end{eqnarray}
as, 
\begin{eqnarray}
~\left[M_{\m\n}^{AT}\right]_{\left[\a\b\right]\left[\g\d\right]}&=&\frac{1}{2}
{\Big(}\left[M^V_{\m\n}\right]_{\a\g}g_{\b\d}+g_{\a\g}
\left[M^V_{\m\n}\right]_{\b\d}
-\left[M_{\m\n}^V\right]_{\a\d}g_{\b\g}-
g_{\a\d}
\left[M_{\m\n}^V\right]_{\b\g}{\Big)}\nonumber\\ 
&=&-2\,\,\mathbf{1}_{\left[\a\b\right]}{}^{\left[\k\sigma \right]}\left[M^V_{\m\n}\right]_{\sigma }
{}^\r\mathbf{1}_{\left[\r\k\right]\left[\g\d\right]}.
\end{eqnarray}

Then, the explicit expression for the Casimir invariant $F$ in (\ref{FCasm}) takes the form,
\begin{eqnarray}
~[F]_{\left[\a\b\right]\left[\g\d\right]}&=&
-\frac{1}{8}{\Big(}\s_{\a\b}\s_{\g\d}- \s_{\g\d}\s_{\a\b}-22\, \mathbf{1}_{\left[\a\b\right]\left[\g\d\right]}{\Big)}.\nonumber\\
 \label{F_AnTeS}
\end{eqnarray}

\subsection{The Lorentz projector on the irreducible 
$\left( \frac{3}{2},0\right)\oplus \left(0,\frac{3}{2} \right) $ sector}
In the
antisymmetric tensor-spinor space under investigation 
there are three Lorentz sectors  of the type $(j_2,j_1)\oplus(j_1,j_2)$,
corresponding to
$(j_2,j_1)$ = $\left(\frac{1}{2},0\right)$, $ \left(\frac{1}{2},1\right)$, and
$\left(\frac{3}{2},0\right)$. The associated representation functions, 
$ w^{(j_1,j_2)}$, are characterized by their $c_{(j_1,j_2)}$ eigenvalues with respect to the $F$
invariant
according to the equations (\ref{FCasm}), and (\ref{Fev}),
\begin{equation}
F w^{(j_2,j_1)}= c_{(j_1,j_2)} w^{(j_2,j_1)}=\frac{1}{2}
(K(K+2)+M^2)w^{(j_1,j_2)},
\label{Finv}
\end{equation}
with
\begin{equation}
K=j_1+j_2, \qquad M=\vert j_1 -j_2 \vert.
\end{equation}
All three eigenvalues are different and given by,
\begin{equation}
c_{\left(\frac{1}{2},0\right)}=\frac{3}{4},\qquad c_{\left(\frac{1}
{2},1\right)}=\frac{11}{4},\qquad
c_{\left(\frac{3}{2},0\right)}=\frac{15}{4}.
\label{tr_org}
\end{equation}
This allows us to  define  the three independent operators,
$\mathcal{Q}^{\left(\frac{1}{2},0\right)}$, and
$\mathcal{Q}^{\left(\frac{1}{2},1\right)}$ as

\begin{eqnarray}
\mathcal{Q}^{(j_1,j_2)}&=&F-c_{(j _1,j
_2)}\mathbf{1},
\end{eqnarray}
which have the properties to remove from  the antisymmetric tensor-spinor in (\ref{tensor_spinor}) 
the dummy   $\left(\frac{1}{2},0\right)\oplus \left(0, \frac{1}{2} \right)$, and
   $\left(\frac{1}{2},1\right)\oplus \left(\frac{1}{2}, 1 \right)$,   companions to $\left(\frac{3}{2},0\right)\oplus\left(0,\frac{3}{2}\right)$. 
Specifically, the Lorentz projector for the irreducible sector
$\left(\frac{3}{2},0\right)\oplus \left(0,\frac{3}{2}\right)$ is cast into the following form,
\begin{equation}
\mathcal{P}^{\left(\frac{3}{2},0\right)}_F=\frac{\mathcal{Q}^{\left(\frac{1}{2},1\right)}
\mathcal{Q}^{\left(\frac{1}{2},0\right)}}{(c_{\left(\frac{3}{2},0\right)}-c_{\left(\frac{1}{2},1\right)})
(c_{\left(\frac{3}{2},0\right)}-c_{\left(\frac{1}{2},0\right)})},
\label{eq30}
\end{equation}
with  $c_{\left(\frac{3}{2},0\right)}$, $c_{\left(\frac{1}{2},1\right)}$, $c_{\left(\frac{1}{2},0\right)}$ from eq.~
(\ref{tr_org}). In this way
we find the following Lorentz projector on $\left(\frac{3}{2},0\right)\oplus \left(0,\frac{3}{2}\right)$,
\begin{eqnarray}
~\left[\mathcal{P}^{\left(\frac{3}{2},0\right)}_F\right]_{\left[\a\b\right]\left[\g\d\right]}&=&\frac{1}
{8}
(\s_{\a\b}\s_{\g\d}+\s_{\g\d}\s_{\a\b})-\frac{1}{12}\s_{\a\b}\s_{\g\d},\nonumber\\
\label{CS_3}
\end{eqnarray}
which satisfies the conditions, 
\begin{eqnarray}
\g^\a \left[\mathcal{P}^{\left(\frac{3}{2},0\right)}_F\right]_{\left[ \a\b\right]\left[ \m\n\right]} &=& 0,\\
\left[\mathcal{P}^{\left(\frac{3}{2},0\right)}_F\right]_{\left[ \a\b\right]\left[\m\n\right]} \g^\m &=& 0.
\end{eqnarray}


\subsection{The spin-$\frac{3}{2}$  wave equation }
We now consider the action of the Lorentz projector in (\ref{eq30})  on the tensor-spinor and its dual and generate in this way the set of twenty four momentum space wave functions, $\left[\phi_\pm^{(\frac{3}{2},0)}\left( {\mathbf p}, \frac{3}{2},\lambda^\prime,\lambda^\prime{}^\prime \right)\right]^{\left[ \alpha\beta  \right]}$,
and $\left[{\widetilde \phi}_\pm^{(\frac{3}{2},0)}\left( {\mathbf p}, \frac{3}{2},\lambda^\prime,\lambda^\prime{}^\prime \right)\right]^{\left[ \alpha\beta  \right]}$, as,
\begin{eqnarray}
\left[\phi^{\left(\frac{3}{2},0\right)}_\pm \left({\mathbf p},\frac{3}{2},\lambda^\prime ,\lambda^\prime{}^\prime \right)\right]^{\left[ \a\b\right]}&=&
\left[\mathcal{P}^{\left(\frac{3}{2},0\right)}_F\right]^{\left[ \alpha \beta\right]}{}{}_{\left[ \g\d\right]}
{\mathcal T}^{\left[ \g\d\right]}_\pm ({\mathbf p},\lambda^\prime, \lambda^\prime{}^\prime ),
\nonumber\\
\left[{\widetilde \phi}^{\left(\frac{3}{2},0\right)}_\pm \left({\mathbf p},\frac{3}{2},\lambda^\prime ,\lambda^\prime{}^\prime \right)\right]^{\left[ \a\b\right]}&=&
\left[\mathcal{P}^{\left(\frac{3}{2},0\right)}_F\right]^{\left[ \alpha \beta\right]}{}{}_{\left[ \g\d\right]}
{\widetilde {\mathcal T}}^{\left[ \g\d\right]}_\pm ({\mathbf p},\lambda^\prime, \lambda^\prime{}^\prime ),\nonumber\\
\lambda^\prime&=&\pm 1,0, \,\, \lambda^\prime{}^\prime=\pm\frac{1}{2}, \nonumber\\
 \lambda&=&\pm\frac{1}{2},\pm \frac{3}{2}.
\label{DRCT}
\end{eqnarray}
After some algebraic manipulations, it can be verified  that these are all $\mathcal{P}^{\left(\frac{3}{2},0\right)}_F$ eigenfunctions, meaning that they all reside in 
$\left(\frac{3}{2}, 0 \right)\oplus \left( 0, \frac{3}{2}\right)$, although only eight of them are linearly independent, as it should be.
Out of them, a set of pure spin-$\frac{3}{2}$  tensor--spinors  of positive and negative parities, henceforth denoted by
$\left[w_\pm ^{\left(\frac{3}{2},0\right)}\left({\mathbf p},\frac{3}{2},\lambda \right)\right]^{\left[ \a\b\right]}$,  can be constructed and
cast into the form of the Lorentz projector   $\left[\mathcal{P}^{\left(\frac{3}{2},0\right)}_F\right]^{\left[\a\b\right]}{}{}_{\left[\m\n\right]}$ 
applied to the anti-symmetric combination, $\left[U_\pm \left({\mathbf p}, \frac{3}{2},\l\right) \right]^{\left[\m\n \right]}$,  
of the four-mo\-men\-tum, $p^\m$, with a four-vector spinor, ${\mathcal U}_\pm ^\n \left({\mathbf p}, \frac{3}{2},\lambda \right)$,
as explained  in eq.~(\ref{ss32ATS}) in the Appendix. In other words,  one can write,
{
\begin{eqnarray}
\left[w_\pm^{\left(\frac{3}{2},0\right)}\left({\mathbf p},\frac{3}{2},\lambda\right)\right]^{\left[ \a\b\right]}&=& 2
\left[\mathcal{P}^{\left(\frac{3}{2},0\right)}_F\right]^{\left[\a\b\right]}{}{}_{\left[\m\n\right]}\left[U_\pm\(\mathbf{p},\frac{3}{2},\l\)\right]^{\left[\m\n\right]}\nonumber\\
&=&\frac{2}{m} \left[\mathcal{P}^{\left(\frac{3}{2},0\right)}_F\right]^{\left[\a\b\right]}{}{}_{\left[\m\n\right]} p^\m {\mathcal U}^\n_\pm\left(\mathbf{p},\frac{3}{2},\l\right),
 \label{spin32ATS}
\end{eqnarray} 
with ${\mathcal U}_\pm ^\nu \left({\mathbf p}, \frac{3}{2},\l\right)$ defined in eqs.~(\ref{fvs324}) in the Appendix.
The factor 2 ensures the normalization of these states to $(\pm 1)$. 
To the amount $\left[w_\pm^{\left(\frac{3}{2},0\right)}\left({\mathbf p},\frac{3}{2},\lambda\right)\right]^{\left[ \a\b\right]}$
are considered on their mass shell, they satisfy the Klein-Gordon equation, as explained
in  (\ref{feq12}), (\ref{masterequations_1}), and  (\ref{Finv}). 
In effect, the  $\left(\frac{3}{2},0 \right)\oplus \left(0,\frac{3}{2} \right)$ degrees of freedom are found to solve the following second order 
equation,                               
\begin{eqnarray}
~ {\Big(} \left[\mathcal{P}^{\left(\frac{3}{2},0\right)}_F\right]^{\left[\a\b\right]}{}{}_{\left[\g\d\right]}
p^2-m^2  \mathbf{1}^{\left[\a\b\right]}{}{}_{\left[\g\d\right]}{\Big)}
\left[w^{\left(\frac{3}{2},0\right)}_\pm \left({\mathbf p},\frac{3}{2},\lambda\right)\right]^{\left[\g\d\right]}
&=&0.\,\,\,
\label{eqATS32}
\end{eqnarray}
In this fashion, the pure spin-$\frac{3}{2}$ degrees of freedom are manifestly generated in the shape of
tensor-spinors. Compared to  the eight-component Weinberg-Joos spin-$\frac{3}{2}$  ``vectors'', listed  among others  in \cite{DelgadoAcosta:2012yc}, 
the  eight tensor-spinors, $\left[w^{\left(\frac{3}{2},0\right)}_\pm \left({\mathbf p},\frac{3}{2},\lambda \right)\right]^{\left[ \a\b\right]}$, open an avenue towards 
efficient tensor calculations of  scattering cross-sec\-tions off $\left(\frac{3}{2},0\right)\oplus \left(0,\frac{3}{2}\right)$ by the help of the symbolic software FeynCalc,
a reason for which we consider the equation (\ref{DRCT}) as  the first achievement of the present study, worth reporting.
The aforementioned tensor spinors do not satisfy the Dirac equation, due to the nonzero commutator,
$\left[ {\mathcal P}_F^{\left(\frac{3}{2},0\right)},p\!\!\!/\right]\not=0$, between the Lorentz-projector and
the Feynman slash, $p\!\!\!/$.  This  is a crucial circumstance for the neat separation of the Rarita-Schwinger-- from the Weinberg-Joos sector in the 
anti-symmetric Lorentz tensor-spinor of second rank. The tensor-spinors are conditioned through, 
\begin{equation}
\gamma^\a\gamma^\b \left[w^{\left(\frac{3}{2},0\right)}_\pm \left({\mathbf p},\frac{3}{2},\lambda \right)\right]_{\left[ \a\b\right]}=0.
\label{TS_GCND}
\end{equation}
We now introduce 
the short-hand $~\left[f^{(\frac{3}{2},0)}{(p)}\right]^{\a\b\m}$ for convenience as
\begin{eqnarray}
~\left[f^{(\frac{3}{2},0)}{(p)}\right]^{\a\b\m}=\frac{2}{m}
\left[\mathcal{P}_F^{(j_1,j_2)}\right]^{\left[\a\b\right]\left[\g\m\right]}p_\g,
\label{f_tensors}
\end{eqnarray}
} with
\begin{eqnarray}
~\left[\overline{f}^{\left(\frac{3}{2},0\right)}(p)\right]^{\a\b}{}_{\m}
\left[f^{\left(\frac{3}{2},0\right)}(p)\right]_{\a\b\n}&=&\frac{1} 
{m^2}  {\Big(}g_{\a\b}g_{\m\n}
-\frac{1}{3}\s_{\a\m}\s_{\b\n}-g_{\a\m}g_{\b\n}{\Big)}p^\a p^\b. \nonumber\\
\label{eq:fintern320}
\end{eqnarray}
This allows us to write the tensor-spinors in (\ref{spin32ATS}) as
\begin{equation}
~\left[
w^{\left(
\frac{3}{2},0\right)}_{\pm}\left(\mathbf{p},\frac{3}{2},\l\right)\right]^{\left[\a\b\right]}=\left[f^{(\frac{3}{2},0)}
(\mathbf{p})\right]^{\a\b}{}{}_\m\left[{\mathcal U}_\pm\left(\mathbf{p},\frac{3}{2},\l\right)\right]^{\m}.
\label{eq:atsvs}
\end{equation}

The conjugate tensor-spinors are then introduced as

\begin{equation}
~\left[\overline{w}^{(\frac{3}{2},0)}_{\pm}(\mathbf{p},\frac{3}{2},\l)\right]_{\left[\a\b\right]}=
\overline{{\mathcal U}}_\pm^\m \left(\mathbf{p},\frac{3}{2},\l\right)
\left[\overline{f}^{\left(\frac{3}{2},0\right)}
(\mathbf{p})\right]_{\a\b}{}{}^\m,
\label{eq:atsvsd}
\end{equation}
were $\left[\overline{f}^{\left(\frac{3}{2},0\right)}
(\mathbf{p})\right]_{\a\b\m}=\g^0\(
\left[f^{\left(\frac{3}{2},0\right)}
(\mathbf{p})\right]_{\a\b\m} \)^\dagger\g^0$. \\
The above spin-$\frac{3}{2}$ tensor-spinors are normalized as
\begin{equation}
\left[\overline{w}^{\left(\frac{3}{2},0\right)}_{\pm}\left(\mathbf{p},\frac{3}{2},\l\right)\right]_{\left[\a\b\right]}
\left[w^{\left(\frac{3}{2},0\right)}_{\pm}\left(\mathbf{p},\frac{3}{2},\l\right)\right]^{\left[\a\b\right]}=\pm 1.
\label{spin_norm}
\end{equation}

\noindent
We now calculate their propagator as, 
\begin{eqnarray}
\label{propsATS}
~\left[S^{\left(\frac{3}{2}, 0\right)}(p)\right]_{\left[\alpha \beta\right] \left[\gamma \delta\right] }=
\frac{\left[\Delta^{\left(\frac{3}{2}, 0\right)}(p)\right]_{\left[\alpha \beta\right] \left[\gamma \delta\right] }}{p^2-
m^2+i\e},
\end{eqnarray}
with 
\begin{eqnarray}
~\left[\D^{\left(\frac{3}{2}, 0\right)}(p)\right]_{\left[\alpha \beta\right]\left[ \gamma \delta \right]}&=&
\frac{p^2}{m^2}\left[\mathcal{P}^{\left(\frac{3}{2},0\right)}_F\right]_{\left[\a\b\right]\left[\g\d\right]}+\frac{\(m^2-p^2\)}{m^2}\mathbf{1}_{\left[\a \b\right] \left[\g \d\right]}.
\end{eqnarray}
\begin{center}
{\footnotesize
\begin{table}
\caption{ \label{table1}
Glossary of the description in momentum space of pure spin-$\frac{3}{2}$ transforming according to the irreducible  
$\left(\frac{3}{2},0\right)\oplus \left(0,\frac{3}{2}\right)$ sector of the 
anti-symmetric Lorentz tensor-spinor of second rank, 
$\left[
{\mathcal T}_\pm \left({\mathbf p}, \lambda^\prime,\lambda^\prime{}^\prime \right)\right]^{\lbrack \mu\nu \rbrack }$. 
Here,  ${\mathcal U}_\pm ^\alpha\left( {\mathbf p},\frac{3}{2},\lambda \right) $ are the four vector-spinor degrees of freedom residing in
the tensor-spinor space and are defined in eqs.~ (\ref{spin32ATS}), (\ref{fvs324}). 
The low case $(\pm)$ index refers to spinors of either positive, $(+)$, or negative, $(-)$, parity,  respectively.}
\begin{tabular}{ll} 
\hline\noalign{\smallskip}
$(j,0)\oplus (0,j)$ & $\left(\frac{3}{2},0\right)\oplus \left(0,\frac{3}{2}\right)\in \left[{\mathcal T}_\pm \left({\mathbf p}, \lambda^\prime,\lambda^\prime{}^\prime \right)\right]^{\lbrack \mu\nu \rbrack }$.  \\
   &\\
\hline\noalign{\smallskip}
Tensor-spinor & $\left[{\mathcal T}_\pm \left({\mathbf p}, \lambda^\prime ,\lambda^\prime{}^\prime \right)\right]^{\lbrack \mu\nu\rbrack }$, \qquad 
eq.~(\ref{TS_MSP}) \\
and its dual:  & $\left[{\widetilde {\mathcal T}}_\pm \left({\mathbf p}, \lambda^\prime ,\lambda^\prime{}^\prime \right)\right]^{\lbrack \mu\nu\rbrack }$ \\
\hline\noalign{\smallskip}
Lorentz projector: &
$\lbrack {\mathcal P}^{\left(\frac{3}{2},0\right)}_F\rbrack_{\lbrack \alpha\beta\rbrack \lbrack \gamma\delta\rbrack }=\frac{1}{8}{\Big(} \sigma_{\alpha\beta}\sigma_{\gamma\delta}+\sigma_{\gamma\delta}\sigma_{\alpha\beta}{\Big)}$\\
&- $\frac{1}{12}\sigma_{\alpha\beta}\sigma_ {\gamma\delta} $,
\qquad eq.~(\ref{CS_3})~\\
   & \\
\hline\noalign{\smallskip}
Primordial spin-$\frac{3}{2}$ & $\left[
\phi_\pm ^{\left(\frac{3}{2},0\right)}\left(\mathbf{p},\frac{3}{2},\lambda^\prime,\lambda^\prime{}^\prime \right)\right]_{\left[ \alpha \beta \right]}=\left[
{\mathcal P}_F^{\left(\frac{3}{2},0\right))}\right]_{\left[ \alpha\beta\right]\left[ \g\d\right]}$\\  &\\
tensor-spinors, &\qquad\qquad  $\times\left[ {\mathcal  T}_\pm ({\mathbf p},\lambda^\prime,\lambda^\prime{}^\prime)\right]^{\left[\g\d\right]} $, \qquad  eqs.~(\ref{DRCT}) \\
 &\\
 and dual tensor-spinors:  & $\left[{\widetilde \phi}_\pm ^{\left(\frac{3}{2},0\right)}\left(\mathbf{p},\frac{3}{2},\lambda^\prime,\lambda^\prime{}^\prime \right)\right]_{\left[ \alpha \beta \right]}=\left[
{\mathcal P}_F^{\left(\frac{3}{2},0\right))}\right]_{\left[ \alpha\beta\right]\left[ \g\d\right]}$ \\
 &$ \times \left[ {\widetilde {\mathcal  T}}_\pm ({\mathbf p},\lambda^\prime,\lambda^\prime{}^\prime)\right]^{\left[\g\d\right]}$\\ 
& \\
\hline\noalign{\smallskip}
Parity & $\left[w_\pm^{\left(\frac{3}{2},0\right)}\left({\mathbf p},\frac{3}{2},\lambda\right)\right]^{\left[ \a\b\right]}
=\frac{2}{m} \left[\mathcal{P}^{\left(\frac{3}{2},0\right)}_F\right]^{\left[\a\b\right]}{}{}_{\left[\m\n\right]}$ \\
representation: & \\
& \qquad $\times p^\m {\mathcal U}^\n_\pm\left(\mathbf{p},\frac{3}{2},\l\right)$, \qquad eq.~(\ref{spin32ATS}),(\ref{ss32ATS})\\
& \\
\hline\noalign{\smallskip}
&\\
Wave equation: & 
$ w_\pm ^{\left(\frac{3}{2},0\right)}\left(\mathbf{p},\frac{3}{2},\lambda\right)_{\lbrack \gamma \delta  \rbrack}$=
$\lbrack {\mathcal P}^{\left(\frac{3}{2},0\right)}_F \rbrack_{\lbrack \alpha\beta\rbrack \lbrack \gamma\delta \rbrack}
\frac{p^2}{m^2}$ \\
&\qquad $\times  \left[ w_\pm ^{\left(\frac{3}{2},0\right)}\left(\mathbf{p},\frac{3}{2},\lambda\right) \right]^{\lbrack \alpha\beta \rbrack}$
\qquad eq.~(\ref{eqATS32}) \\
   &\\
\hline\noalign{\smallskip}
Electromagnetic  ~&
$j_{\mu}^{\left(\frac{3}{2},0\right)}(\mathbf{p}',\l',\mathbf{p},\l)=
\frac{38}{9}\,e\,
\overline{{\mathcal U}}^\a_+
\left(\mathbf{p}',\frac{3}{2},\l'\right)$\\
current: &$ {\Big(}
(p'+p)_\m g_{\a\b}-m g_{\a\b}\g_\m-(p'_\b g_{\a\m}+p_\a
g_{\b\m})$\\
&$+\frac{20}{38 m}(p_\alpha p^\prime_\beta -p^\prime\cdot p g_{\alpha \beta})\gamma_\mu 
{\Big)}{\mathcal U}^\beta_+\left(\mathbf{p},\frac{3}{2},\lambda\right)$,\\
~& \\
   & \,\, eq.~(\ref{elmcr})\\
&\\
\hline\noalign{\smallskip}\hline
\end{tabular}
\end{table}
}
\end{center}
Second order theories present the notorious  problem that the propagators are of unspecified spatial parities. 
{}For bosons this circumstance does not present an obstacle in so far as bosons and anti-bosons are
of equal parities. Such is due to the commutativity of the parity and the charge conjugation operators.
A discussion on this issue can be found  around the equation (2.34) in the reference ~\cite{DelgadoAcosta:2012yc}.
{}For massive charged fermions, however,  for which  particles and anti-particles are of opposite parities, the  problem is more serious
and can affect the quantization procedure. However, we expect that the method would allow for quantization of 
 Majorana fermions (if they were to exist) and  massless particles.
Moreover, at the level of  scattering processes, the problem of parity distinction of charged massive particles still can be attended  by constructing the amplitudes for states of a fixed parity and 
then inserting into  the squared amplitudes the relevant parity projector, a path that we take in the evaluation of the process 
of Compton scattering as presented in subsection 5 of the next section. The method is highlighted in table 1.

\section{The coupling of the  $\left( \frac{3}{2},0\right)\oplus \left( 0, \frac{3}{2}\right)$ 
sector of the anti-symmetric tensor spinor to the electromagnetic field}\label{sec4}

As a next step after having designed  the appropriate free equations of motion
in (\ref{masterequations_1}), (\ref{eqATS32}),
we  introduce the electromagnetic interaction.
We here  confine to minimal gauging, i.e.  the coupling is found by replacing
ordinary by covariant derivatives
according to,
\begin{equation}
\partial^\mu\longrightarrow D^\mu=\pd^\mu+i e A^\mu,
\label{gauge_simtrnsf}
\end{equation}
where  $e$ is the electric
charge of the particle.

\subsection{The gauge procedure and the number of spin degrees of freedom  upon gauging}\label{sec4a}

In order to obtain the gauged equations, we first pass the states from momentum  to 
position space using the standard quantization prescription, 
$[w ^{\left(\frac{3}{2}, 0\right)} \left({\mathbf p},\frac{3}{2},\lambda \right) ]^{\left[ \a\b\right]}e^{\mp i\,x\cdot p}$
$\longrightarrow$
$[\y^{\left(\frac{3}{2},0\right)} (x,\l)]^{\left[ \a\b\right]}$, and then write the momenta in operator form yielding,
\begin{equation}
[\mathcal{P}^{\left(\frac{3}{2},0\right)}_F]_{[\a\b]}{}{}^{[\gamma\delta]}\partial^2
[\y ^{\left(\frac{3}{2},0\right)}(x,\l)]^{\lbrack \a\b\rbrack} = -m^2 [\y^{\left(\frac{3}{2},0\right)} (x,\l)]^{\lbrack \g\d\rbrack}.
\label{fdEqs32_prl}
\end{equation}
We now define at the free particle level a new tensor,
$[\G^{\left(\frac{3}{2}, 0\right)}_{\m\n}]^{ \left[\gamma \delta\right]  }{}{}_{\left[ \a\b\right]}$, as
\begin{eqnarray}
~[\G^{\left(\frac{3}{2},0\right)}_{\m\n}]_{[\a\b][\g\d]}\partial ^\m \partial ^\n=
[\mathcal{P}^{\left(\frac{3}{2},0\right)}_F]_{[\a\b][\gamma\delta]}\partial^2,
\label{G_3}
\end{eqnarray}
and then cast the gauged  eq.~(\ref{fdEqs32_prl}) as,
\begin{equation}
[\G^{\left(\frac{3}{2},0\right)}_{\m\n}]_{[\a\b][\g\d]}D^\m D^\n
[\Psi^{\left(\frac{3}{2},0\right)} (x)]^{\lbrack \a\b\rbrack} = -m^2 [\Psi^{\left(\frac{3}{2},0\right)} (x )]^{\lbrack \g\d\rbrack},
\label{fdEqs32}
\end{equation}
where we denoted by $\left[\Psi^{\left(\frac{3}{2},0\right)} (x)\right]^{\left[ \a\b \right]}$  the new gauged solutions. 
In order to guarantee that the gauged  solutions  continue 
being eigenstates of the Lorentz projectors and thereby to ensure equality  of the number of the 
degrees of freedom before and after gaugung, the 
$[\G^{\left(\frac{3}{2},0\right)}_{\m\n}]_{[\a\b][\g\d]}$ tensor has to satisfy,
\begin{equation}
\left[ {\mathcal P}_F^{\left(\frac{3}{2},0\right)}\right]_{[\a  \b]}{}{}^{\left[\sigma \rho\right]}\left[ \Gamma^{\left(\frac{3}{2},0\right)}_{\mu\nu}\right]_{\left[\sigma  \rho\right][\gamma\delta]}=
\left[ \Gamma^{\left(\frac{3}{2},0\right)}_{\mu\nu}\right]_{[\alpha  \beta][\gamma\delta]}.\label{G32tensor}
\end{equation}
Once  the validity of the equation  (\ref{G32tensor}) has been ensured, the representations space and therefore the spin after gauging continues being same as before.
It should be noticed that the $\Gamma^{\left(\frac{3}{2},0\right)}_{\mu\nu}$ tensor in 
(\ref{G_3}) is determined at the free particle level   modulo additive terms leading to vanishing $\left[\partial ^\mu,\partial ^\nu\right]$
commutators prior gauging, which upon gauging give rise  to the electromagnetic  field strength tensor, $F_{\m\nu}$ and thereby to non-vanishing contributions to the gauged equations.
As first discussed in \cite{Napsuciale:2006wr}, exploiting this freedom  could be of help  in achieving 
causality of particle propagation, and/or  unitarity of scattering amplitudes in the ultraviolet.

\subsection{Causality of  $\left( \frac{3}{2},0\right)\oplus \left(0,\frac{3}{2} \right)$ propagation upon gauging}

Using  the aforementioned freedom we chose  $\left[\G^{\left(\frac{3}{2}, 0\right)}_{\m\n}\right]^{ \left[\a \b\right]  }{}{}_{ \left[\g\d\right]}$ in such a way that, 
\begin{eqnarray}
~\left[
\G^{\left(\frac{3}{2}, 0\right)}{}^\m{}{}_\n
\right]^{\left[ \a \b\right]  }{}{}_{ \left[\g\d\right]}=4\left[\mathcal{P}^{\left(\frac{3}{2},0\right)}_F\right]^{\left[\a\b\right]
\left[{ \sigma} \m\right]}\
\left[\mathcal{P}^{\left(\frac{3}{2},0\right)}_F\right]_{\left[\sigma \n\right]\left[\g\d\right]}&&\nonumber\\
=\frac{1}{2}{\Big(}
{\s}^{\alpha\beta}{ \s}^{\sigma\mu}{\mathbf \s}_{\sigma\nu}{\s}_{\gamma\delta}+
{ \s}^{\sigma\mu}{\s}^{\alpha\beta}{\s}_{\gamma\delta}{\s}_{\sigma\nu}&&\nonumber\\
-3{ \s}^{\alpha\beta}{\s}^{\sigma\mu}{ \s}_{\gamma\delta}
{\s}_{\sigma\nu}-3{\s}^{\sigma\mu}{ \s}^{\alpha\beta}{\s}_{\sigma\nu}
{\s}_{\gamma\delta}
{\Big)}&&\nonumber\\
+4{\s}^{\sigma\mu}{ \s}^{\alpha\beta}{\s}_{\sigma\nu}{ \s}_{\gamma\delta},&&\nonumber\\
\label{gamma-s32}
\end{eqnarray}
is satisfied, thus ensuring validity of (\ref{G_3}) and (\ref{G32tensor}). 
This is the precise tensor that enters the  gauged equation for the  $\left( \frac{3}{2},0\right)\oplus \left(0,\frac{3}{2} \right)$ sector in
(\ref{fdEqs32}).\\
   
\noindent
By virtue of (\ref{G32tensor}), on expects the solutions to \eqref{fdEqs32} to continue behaving as 
$\left( \frac{3}{2},0\right)\oplus\left( 0,\frac{3}{2} \right)$ states
and depending  on only eight degrees of freedom. In order to see this, it is convenient to become aware of the fact that that upon accounting for the Dirac label,
the equation ~(\ref{fdEqs32}) in reality stands for a $(24\times 24)$  dimensional matrix equation.
It is straightforward to cross-check that the matrix in (\ref{fdEqs32}) in combination with (\ref{gamma-s32}) can be block-diagonalized 
with one of the blocks being precisely the expected $(8\times 8)$  dimensional one.
The corresponding eight-dimensional  gauged solutions, $\Psi^{\left(\frac{3}{2},0\right)} (x)$, can be now written as linear combinations within any  basis spanning  
the $\left(\frac{3}{2},0\right)\oplus\left(0,\frac{3}{2}\right)$  representation space.  
In particular, if we choose the orthogonal set of the rest-frame states given by (\ref{eq:atsvs}), 
$\Psi^{\left( \frac{3}{2},0\right)} (x)$ can be decomposed according to,   
\begin{equation}
\Psi ^{\left(\frac{3}{2},0\right)}(x )=\sum_{\l^\prime{}^\prime }{\Big(}  a_+(x)_{\l^\prime{}^\prime  }w^{(\frac{3}{2},0)}_{+}({\mathbf 0},\frac{3}{2},\l^\prime{}^\prime  )
-a_-(x)_{\l^\prime{}^\prime  } w^{(\frac{3}{2},0)}_{-}(\mathbf{0},\frac{3}{2},\l^\prime{}^\prime  )
{\Big)}.
 \label{sollc}
\end{equation}
As a reminder, the reduction to eight degrees of freedom is possible  by virtue of the condition in (\ref{G32tensor})
imposed by the  Lorentz projector on the  $\left[ \Gamma^{\left(\frac{3}{2},0\right)}_{\mu\nu}\right]_{[\alpha  \beta][\gamma\delta]}$ tensor which amounted to (\ref{gamma-s32}).
Below we show that the expansion of the gauged solutions in (\ref{sollc}) significantly  facilitates the proof of their causal  propagation within an 
electromagnetic environment.\\   

\noindent
The causality and hyperbolicity of the equations of motion of order $\leq 2$
in the derivatives
are tested using the Courant-Hilbert criterion, which requires us to
calculate the so called ``characteristic determinant'' \cite{Velo:1970ur} of the gauged equation.
The latter is  found by replacing the highest order  derivatives  by the components of
the vectors $n^\m$, interpreted as the normals to the characteristic surfaces,
and which characterize  the propagation of the (classical)  wave fronts of the gauged
equation.
If the vanishing  of the
characteristic determinant demands to have a real-valued time-like component
$n^0$, then the equation
is hyperbolic. If this determinant nullifies for $n^\m n_\m=0$, then the
equation is in addition  causal \cite{Velo:1970ur}.
\noindent
The $a_\pm(x)_{\l^\prime }$ coefficients are calculated upon substitution of  (\ref{sollc}) in (\ref{fdEqs32}), and  
making use of the normalization of the basis tensors in  (\ref{spin_norm}) as,
\begin{equation}
\left[\overline{w}^{\left(\frac{3}{2},0\right)}_{\pm}(\mathbf{0},\frac{3}{2},\l^\prime )\right]_{\left[\a\b\right]}
\left[\G^{\left(\frac{3}{2}, 0\right)}_{\m\nu}\right]^{ \left[\a \b\right]  }{}{}_{\left[ \g\d\right]}D^\m D^\n\left[\Psi^{\left(\frac{3}{2},0\right)}(x )\right]^{\left[\g\d\right]}= - m^2 a_\pm(x)_{ \l^\prime },
\end{equation}
meaning that they satisfy the equation,

\begin{eqnarray}
\left[\overline{w}^{\left(\frac{3}{2},0\right)}_{\pm}\left(\mathbf{0},\frac{3}{2},\l^\prime \right)\right]_{\left[\a\b\right]}
&&\left[\G^{\left(\frac{3}{2}, 0\right)}_{\m\nu}\right]^{ \left[\a \b\right]  }{}{}_{\left[ \g\d\right]}D^\m D^\n
\sum_{\l^\prime {}^\prime }a_\pm (x)_{ \l^\prime{}^\prime }\left[w^{\left(\frac{3}{2},0\right)}_{\pm}\left({\mathbf 0},\frac{3}{2},\l^\prime{}^\prime 
\right)\right]^{\left[\g\d\right]}\nonumber\\
&=&- m^2 a_\pm(x)_{ \l^\prime },\label{eqfora}\\
D^\mu D^\nu&=&\partial^\mu\partial^\nu +ie(\partial^\m A^\n) +ieA^\nu\partial^\mu 
+ieA^\mu\partial^\nu -e^2A^\mu A^\nu.\label{GD_PRDCT}
\end{eqnarray}
The equation \eqref{eqfora} represents a system of eight partial differential equation for the eight  coefficients 
$a_\pm(x)_{\l^\prime{}^\prime}$,  because the polarization label $\l^\prime{}^\prime  $ runs over  the four allowed values, 
$\lambda^\prime{}^\prime   =\pm \frac{3}{2}, \pm \frac{1}{2}$, and is counted twice because of the two parities (the lower case $\pm $ indexes), 
of the basis  tensor-spinors, $w^{\left(\frac{3}{2},0\right)}_\pm \left({\mathbf 0}, \frac{3}{2},\l^\prime{}^\prime \right)$. 
The elements of the characteristic  matrix are proportional to $n^2$, as can been seen from replacing the principal part, $\partial^\mu\partial^\nu$  of
$D^\mu D^\nu$ in (\ref{GD_PRDCT}) by $n^\mu n^\mu$  yielding,
\begin{eqnarray}
\left[\overline{w}^{\left(\frac{3}{2},0\right)}_{\pm}\left(\mathbf{0},\frac{3}{2},\l^\prime \right)\right]_{\left[\a\b\right]}
\left[\G^{\left(\frac{3}{2}, 0\right)}_{\m\nu}\right]^{ \left[\a \b\right]  }{}{}_{\left[ \g\d\right]}n^\m n^\n&&\nonumber\\
\times \eta_t\left[w^{\left(\frac{3}{2},0\right)}_{\pm} \left({\mathbf 0},
\frac{3}{2},\l^\prime{}^\prime  \right)\right]^{\left[\g\d\right]}=
\left[\overline{w}^{\left(\frac{3}{2},0\right)}_{\pm}\left(\mathbf{0},\frac{3}{2},\l^\prime \right)\right]_{\left[\a\b\right]}\nonumber\\
\times n^2\left[\mathcal{P}_F^{\left(\frac{3}{2}, 0\right)}\right]^{ \left[\a \b\right]  }{}{}_{\left[ \g\d\right]}\eta_t
\left[w^{\left(\frac{3}{2},0\right)}_{\pm}\left({\mathbf 0},\frac{3}{2},\l^\prime{}^\prime  \right)\right]^{\left[\g\d\right]}&&\nonumber\\ 
=n^2
\left[\overline{w}^{\left(\frac{3}{2},0\right)}_\pm \left({\mathbf 0},\frac{3}{2},\l^\prime\right)\right]_{\left[\a\b\right]}
\eta_t\left[w^{\left(\frac{3}{2},0\right)}_{\pm}\left({\mathbf 0},\frac{3}{2},\l^\prime{}^\prime  \right)\right]^{\left[\a\b\right]}&&\nonumber\\
=n^2 \delta _{\l^\prime \l^\prime {}^\prime }\delta_{\pm \pm},&\quad &\nonumber\\
 t=\pm, \, \eta_+=-\eta_-=1.&&\nonumber\\
\end{eqnarray}
The characteristic  matrix in question is therefore diagonal and each element equals  $n^2$, correspondingly, its  determinant is
\begin{equation}
\mathcal{D}^{\left(\frac{3}{2},0\right)}(n)=(n^2)^8.
\end{equation}

Nullifying the latter amounts to the condition $n^2=n^\mu n_\mu=0$, which  guarantees  causal
propagation within the electromagnetic environment.


\subsection{The Lagrangian for  $\left(\frac{3}{2},0 \right)\oplus \left(0,\frac{3}{2} \right)$ as part of  the anti-symmetric tensor spinor }

\begin{figure}
\includegraphics{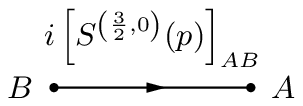}
\caption{\label{propfig} Feynman rule for the spin-$\frac{3}{2}$ propagator in (\ref{propsATS}) of
a particle transforming in the
$\left(\frac{3}{2},0\right)\oplus\left(0, \frac{3}{2}\right)$ sector of the anti-symmetric Lorentz tensor-spinor of second rank.
All the subsequent figures refer to this case.}

\includegraphics{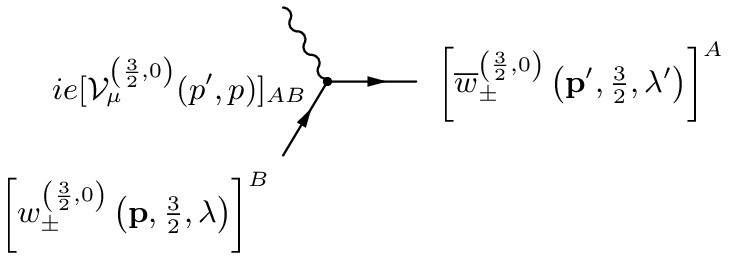}
\caption{\label{regla1fig} Feynman rule for the one-photon vertex in (\ref{vert1}).}

\includegraphics{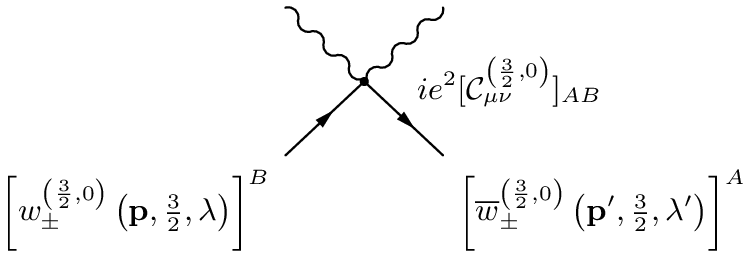}
\caption{\label{regla2fig} Feynman rule for the two-photon contact vertex
in (\ref{vert2}).}
\end{figure}

The free equations of motion (\ref{fdEqs32}) can now
be  derived from the following Lagrangian:
\begin{eqnarray}
{\mathcal L}^{\(\frac{3}{2},0\)}_\textup{free}&=&{\Big(}\pd^\m
[\overline{\y}^{\(\frac{3}{2},0\)}]^A{\Big)}
[\G_{\m\n}^{\(\frac{3}{2},0\)}]_{AB}\pd^\n[\y^{\(\frac{3}{2},0\)}]^B-m^2[\overline{\y}^{\(\frac{3}{2},0\)}]^A[\y^{\(\frac{3}{2},0\)}]_A,\nonumber\\
A&=& \left[\mu\nu\right],\quad B=\left[ \g\d\right],
\label{Lagr}
\end{eqnarray}
where we suppressed the arguments for the sake of simplifying notations.
The gauged Lagrangian then emerges as,
\begin{eqnarray}
{\mathcal L}^{\(\frac{3}{2},0\)}&=&\left(D^\ast\, ^\m [
\overline{\Psi}^{\(\frac{3}{2},0\)}]^A\right)
[\G_{\m\n}^{\(\frac{3}{2},0\)}]_{AB}D^\n[\Psi^{\(\frac{3}{2},0\)}]^B
-m^2[\overline{\Psi}^{\(\frac{3}{2},0\)}]^A[\Psi^{\(\frac{3}{2},0\)}]_A,\nonumber\\
\end{eqnarray}
and its decomposition into free and interacting parts reads,
\begin{eqnarray}\label{current-s32}
{\mathcal L}^{\(\frac{3}{2},0\)}&=&{\mathcal L}^{\(\frac{3}{2},0\)}_\textup{free}+
{\mathcal L}^{\(\frac{3}{2},0\)}_\textup{int},\\
{\mathcal L}^{\(\frac{3}{2},0\)}_\textup{int}&=&-
j_\m^{\(\frac{3}{2},0\)}A^\m+k_{\m\n}^{\(\frac{3}{2},0\)}A^\m A^\n.
\end{eqnarray}
Back to momentum space, we extract the following vector and tensor  bi-linear forms, 
\begin{eqnarray}
j_\m^{\(\frac{3}{2},0\)}({\mathbf p},\l, {\mathbf p}^\prime,\l^\prime)&=&e
\left[\overline{w}_\pm^{\(\frac{3}{2},0\)}\left(\mathbf{p}',\frac{3}{2} ,\l'\right)\right]^A[
\mathcal{V}^{\left(\frac{3}{2},0\right)}_\m(p',p)]_{AB} \left[w_\pm^{\left(\frac{3}{2},0\right)}\left(\mathbf{p},\frac{3}{2},\l\right)\right]^B,\nonumber\\
\label{ECRRNT}\\
k_{\m\n}^{\left(\frac{3}{2},0\right)}({\mathbf p},\l, {\mathbf p}^\prime,
\l^\prime)&=&e\left[\overline{w}_\pm ^{\left(\frac{3}{2},0\right)}\left(\mathbf{p}',\frac{3}{2},\l'\right)\right]^A
 [\mathcal{C}^{\(\frac{3}{2},0\)}_{\m\n}]_{AB}\left[w_\pm^{\(\frac{3}{2},0\)}\left(\mathbf{p},\frac{3}{2},\l\right)\right]^B,
\label{ktnsr}
\end{eqnarray}
the first standing for the electromagnetic current.
The vertexes are given as,
\begin{eqnarray}
~[\mathcal{V}^{\(\frac{3}{2},0\)}_\m(p',p)]_{AB}&=&
[\G^{\(\frac{3}{2},0\)}_{\n\m}]_{AB}p'^\n+
[\G^{\(\frac{3}{2},0\)}_{\m\n}]_{AB}p^\n,\label{vert1}\\
~[\mathcal{C}^{\(\frac{3}{2},0\)}_{\m\n}]_{AB}&=&\frac{1}{2}\
([\G^{\(\frac{3}{2},0\)}_{\m\n}]_{AB}+[\G^{\(\frac{3}{2},0\)}_{\n\m}]_{AB} )\label{vert2}.
\end{eqnarray}
The Feynman rules following  from this Lagrangian are depicted in Figs.
\ref{propfig}, \ref{regla1fig}, \ref{regla2fig}.
It is not difficult to verify that the one-photon vertex obeys 
\begin{equation}
(p'-p)^\m [{\mathcal V}_\m^{\(\frac{3}{2},0\)}(p',p)]_{AB}=[S^{\(\frac{3}{2},0\)}
(p')]_{AB}-[S^{\(\frac{3}{2},0\)}(p)]_{AB},
\label{WT_Id}
\end{equation}
which is the Ward-Takahashi identity, as it should be.

\subsection{Electromagnetic multipole moments }



We now calculate the electromagnetic  multipole moments of  a particle transforming in the
single spin- $\frac{3}{2}$ irreducible Weinberg-Joos sector of the
antisymmetric tensor-spinor of interest from the 
 current in (\ref{ECRRNT}), taken between positive parities. This current can be further simplified in taking advantage of
the equations (\ref{spin32ATS}) from above, of eqs.~ (\ref{ss32ATSprl})-(\ref{ss32ATS})  in the Appendix, and of the mass shell condition,
yielding,
\begin{eqnarray}
j_{\mu}^{\left(\frac{3}{2},0\right)}
(\mathbf{p}',\l',\mathbf{p},\l)&=&
\frac{38}{9}\,e\,
\overline{{\mathcal U}}^\a_+
{\Big(}\mathbf{p}',\frac{3}{2},\l'{\Big)}\nonumber\\
&&{\Big(}(p'+p)_\m g_{\a\b}
-m g_{\a\b}\g_\m -(p'_\b g_{\a\m}+p_\a g_{\b\m})\nonumber\\
&+&\frac{20}{38 m}( p_\a p'_\b-p'\cdot p \,g_{\a\b})\g_\m {\Big)}{\mathcal U}^\b _+{\Big(}\mathbf{p},\frac{3}{2},\l{\Big)}.
\label{elmcr}
\end{eqnarray}
The procedure of extracting the multipole moments from known currents is 
well established  and will not be highlighted here (see for example \cite{Lorce:2009bs},\cite{DelgadoAcosta:2012yc} and references therein
for technical details), and amounts to
\begin{eqnarray}
~[Q_E^0(\l)]&=&e,\label{QE0}\\
~[Q_M^1(\l)]&=&\frac{2}{3}\(\frac{e}{2m}\)\langle M_{12}\rangle,\label{QM1}\\
~[Q_E^2(\l)]&=&-\frac{1}{3}\(\frac{e}{m^2}\)\langle {\mathcal A}\rangle,\label{QE2}\\
~[Q_M^3(\l)]&=&-2\(\frac{e}{2m^3}\)\langle {\mathcal B}\rangle,\label{QM3}
\end{eqnarray}
with
\begin{eqnarray}
{\mathcal A}=3M_{12}^2 -{\mathbf J}^2, &\quad& {\mathcal B}=M_{12}\left(15 M_{12}^2 -\frac{41}{5}{\mathbf J}^2, \right), \nonumber\\
{\mathbf J}^2&=&M_{12}^2+M_{13}^2 +M_{23}^2.
\end{eqnarray}
The  expressions in (\ref{QE0})--(\ref{QM3})  fully coincide in form with the corresponding predictions by the  
Weinberg-Joos formalism reported in
\cite{DelgadoAcosta:2012yc},  
where  the calculation has been carried out while treating the states under
consideration as eight-component spinors.
The only difference concerns  the value of the  gyromagnetic ratio which in the present work comes out fixed
to the inverse of the spin, $g=\frac{2}{3}$, and in accord with Belinfante's conjecture, while  in
\cite{DelgadoAcosta:2012yc}, a method exclusively based on the Poincar\'e covariant spin-projector alone,
$g$ had remained unspecified according to,
\begin{eqnarray}
~[Q_E^0(\l)]_{TS}&=&e,\\
~[Q_M^1(\l)]_{TS}&=&g\(\frac{e}{2m}\)\langle M_{12}\rangle,\\
~[Q_E^2(\l)]_{TS}&=&-(1-g)\(\frac{e}{m^2}\)\langle {\mathcal A}\rangle,\\
~[Q_M^3(\l)]_{TS}&=&-3 g\(\frac{e}{2m^3}\)\langle {\mathcal B}\rangle.
\end{eqnarray}
We conclude that the anti-symmetric tensor-spinor provides a  description of particles of
spin-$\frac{3}{2}$ transforming as $\left(\frac{3}{2},0 \right)\oplus \left(0,\frac{3}{2} \right)$  that is equivalent to the Weinberg-Joos formalism. The great advantage of the tensor basis employed here that it
allows to comfortably carry out more complicated calculations such a Compton scattering, the subject of the next section.


\subsection{Compton scattering off spin-$\frac{3}{2}$ particles in $\left( \frac{3}{2},0\right)\oplus \left( 0, \frac{3}{2}\right)$}

\begin{figure}
\includegraphics{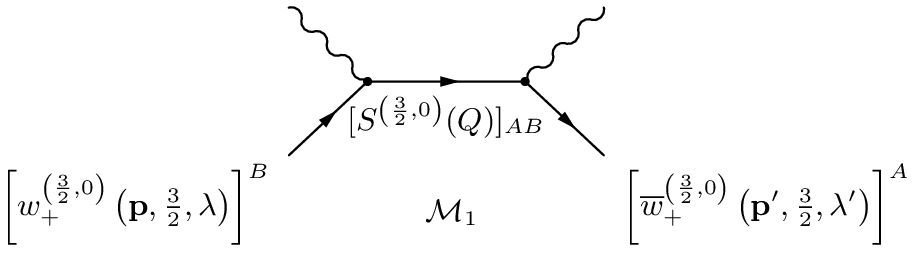}
\caption{\label{mafig} Diagram for the direct-scattering contribution
(\ref{maj1}) to the Compton scattering amplitude (\ref{camp}).}
\includegraphics{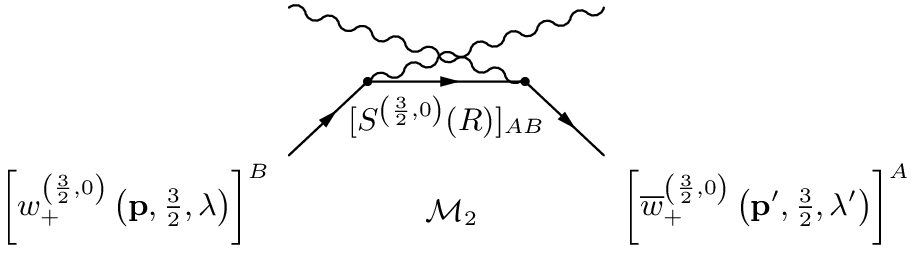}
\caption{\label{mbfig} Diagram for the exchange-scattering contribution
(\ref{mbj1}) to the Compton scattering amplitude (\ref{camp}).}
\includegraphics{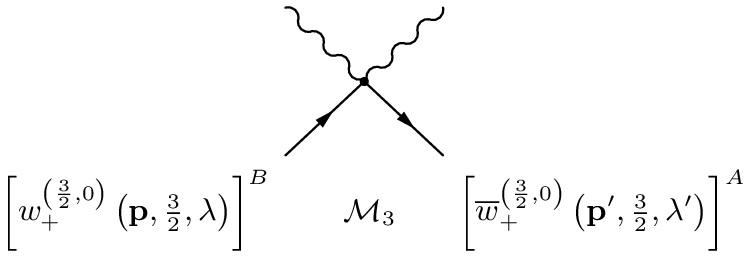}
\caption{\label{mcfig} Diagram for the point-scattering contribution
(\ref{mcj1}) to the Compton scattering amplitude (\ref{camp}).}
\end{figure}

The tree-level Compton scattering amplitude  contains contributions from
the three different channels displayed in the  Figs. \ref{mafig}, \ref{mbfig}, \ref{mcfig}.
In denoting by $p$ and $p^\prime $  in turn the four-momenta of the incident and scattered
single spin-$\frac{3}{2}$
target particles, while by  $q$ and $q^\prime $   the four-momenta of the incident
and scattered photons, respectively,  the amplitude takes the following form \cite{Bjorken},
\begin{eqnarray}\label{camp}
\mathcal{M}&=&\mathcal{M}_1+\mathcal{M}_2+\mathcal{M}_3.
\end{eqnarray}
In the following we shall evaluate the process off positive parity states, in which case 
one has,  
\begin{eqnarray}
i{\mathcal M}_1=e^2\left[
{\overline w}_+^{\left(\frac{3}{2},0\right)}
\left({\mathbf p}^\prime ,\frac{3}{2},\l^\prime \right)
\right]^{A}
&&\left[
{\mathcal U}_{\m\n}(p^\prime ,Q,p)
\right]_{AB}\nonumber\\
\left[
w_+^{\left(\frac{3}{2},0\right)}\left({\mathbf p},\frac{3}{2},\l\right)
\right]^{B}&&
\left[\e^\m({\mathbf q}^\prime ,\ell^\prime )\right]^* \e^\n({\mathbf q},\ell),\nonumber\\
\label{maj1}\\
i{\mathcal M}_2 =e^2\left[ 
{\overline w}_+^{\left(\frac{3}{2},0\right)}
\left({\mathbf p}^\prime ,\frac{3}{2},\l'\right)
\right]^{A}&&
\left[{\mathcal U}_{\n\m}(p',R,p)\right]_{AB}\nonumber\\
\left[ w_+^{\left(\frac{3}{2},0\right)}\left({\mathbf p},\frac{3}
{2},\l \right)\right]^{B}&&[\e^\m(\mathbf{q}', \ell^\prime )]^*
\e^\n({\mathbf q},\ell),\nonumber\\
\label{mbj1}\\
-i{\mathcal M}_3=e^2\left[
{\overline w}_+^{\left(\frac{3}{2},0\right)}\left({\mathbf p}^\prime ,\frac{3}
{2},\l'\right)\right]^{A}&&\left[{\mathcal X}_{\m\n}\right]_{AB}\nonumber\\
\left[ w_+^{\left(\frac{3}{2},0\right)}\left({\mathbf p},\frac{3}{2},\l\right)\right]
^{B}&&\left[\e^\m({\mathbf q}^\prime ,\ell^\prime )\right]^*\e^\n({\mathbf q},\ell)\nonumber\\
\label{mcj1},
\end{eqnarray}
with $Q=p+q^\prime =p^\prime+q^\prime $ and $R=p^\prime -q=p-q^\prime $. Furthermore, we define the short-hands, 
\begin{eqnarray}
~[\mathcal{U}_{\m\n}(p^\prime ,Q,p)]_{AB}&=&\left[\mathcal{V}^{\left(\frac{3}
{2},0\right)}_{\m}
(p',Q)\right]_{AC}
~\left[S^{\left(\frac{3}{2},0\right)}
(Q)\right]^{CD}\nonumber\\
&&\left[\mathcal{V}^{\left(\frac{3}{2},0\right)}_\n(Q,p)\right]_{DB},\\
~[\mathcal{X}_{\m\n}]_{AB}&=&\left[\mathcal{C}_{\m\n}^{\left(\frac{3}
{2},0\right)}+\mathcal{C}_{\n\m}^{\left( \frac{3}
{2},0\right)}\right]_{AB},
\end{eqnarray}
with ${\mathcal V}_\mu^{\left(\frac{3}{2},0\right)}$ and $C_{\mu\nu}^{\left(\frac{3}{2},0\right)}$ defined in (\ref{vert1}) and (\ref{vert2}), respectively, and the indexes $A$,$B$ being introduced  in (\ref{Lagr}).
The gauge invariance of this amplitude is ensured by the Ward-Takahashi
identity in (\ref{WT_Id}). The averaged squared amplitude then reads,
\begin{eqnarray}\label{m2av}
\overline{\left\vert\mathcal{M}\right\vert^2}&=&\frac{1}
{8}\sum_{\l,\l',\ell,\ell'}
\mathcal{M}[{\mathcal M}]^\dagger\\
&=&\frac{1}{8}Tr\left[[\mathcal{M}_{\m\n}(p',Q,R,p)]_{AB}
[{\mathcal M}^{\n\m}(p,R,Q,p')]^{AB}\right],  \label{eq:asa}\nonumber\\
\end{eqnarray}
and the corresponding expression contains  the projector on positive parity states, 
${\mathbf P}_+^{\left(\frac{3}{2}, 0\right)}({\mathbf p}^\prime)$, as
\begin{eqnarray}
~[{\mathcal M}_{\m\n}(p',Q,R,p)]_{AB}&=&
e^2[{\mathbf P}^{\left(\frac{3}{2},0\right)}_+
({\mathbf p}')]_{A}{}^C[\mathbf{U}_{\m\n}(p',Q,R,p)]_{CB},\nonumber\\
~[{\mathbf U}_{\m\n}(p',Q,R,p)]_{CB}&=&{\Big(}{\mathcal U}_{\m\n}(p',Q,p)+{\mathcal U}_{\n\m}(p',R,p)-{{\mathcal X}}_{\m\n}{\big)}_{CB}.
\end{eqnarray}
The projector over spin-$\frac{3}{2}$ states of positive parity has same structure as the one given by \cite{VanNieuwenhuizen:1981ae},
\begin{eqnarray}
&&[{\mathbf P}^{\left(\frac{3}{2},0\right)}_+({\mathbf p})]_{AB}=
\sum_{\l}\left[ w_+^{\left(\frac{3}{2},0\right)}\left({\mathbf p},\frac{3}
{2},\l\right)\right]_{A}
\left[\overline{w}_+^{\left(\frac{3}{2},0\right)}
\left({\mathbf p},\frac{3}{2},\l\right)\right]_{B}.
\label{phpr}
\end{eqnarray}

\begin{figure}
\includegraphics[scale=0.65]{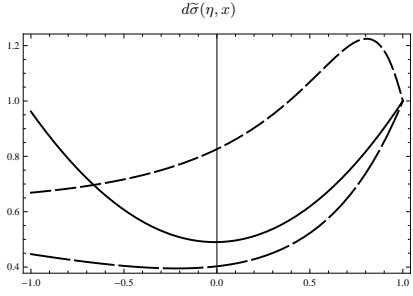}
\caption{\label{dsfig} The differential cross section, $d\widetilde{\s}(\eta ,x)$, as a
function of $x=\cos\q$. The solid curve represents the classical limit at
$\eta =\omega /m=0$, the long dashed line corresponds to an energy comparable to the mass
of the particle, $\eta= 1$, while the short dashed curve corresponds to $\eta=2.5$.
This cross section increases with energy, except in the forward
direction, $x=1$, where it approaches  $d\widetilde{\s}(\eta ,1)=1$.}
\end{figure}

We furthermore used
\begin{eqnarray}
\sum_{\ell}\e^\m(\mathbf{q},\ell)[\e^\n(\mathbf{q},\ell)]^{*}&=&-
g^{\m\n}.
\end{eqnarray}
The pure spin-$\frac{3}{2}$  projector, $[\mathbf{P}^{\left(\frac{3}{2},0\right)}_+(\mathbf{p})]_{AB}$ can 
equivalently be rewritten to
\begin{eqnarray}
~\left[\mathbf{P}_{+}^{\left(\frac{3}{2},0\right)}
(\mathbf{p})\right]_{AB}
&=&\left[f^{\left(\frac{3}{2},0\right)}(\mathbf{p})\right]_{A}{}^{\m}\(\frac{-
{p}\!\!\!/+m}{2m}\) \left[\overline{f}^{\left(\frac{3}{2},0\right)}(\mathbf{p})\right]_{B\m},
\end{eqnarray}
with $f^{\left( \frac{3}{2},0\right)}({\mathbf p})$ from (\ref{f_tensors}). 
The contractions indicated in (\ref{eq:asa}) are comfortably evaluated with the aid
of the  FeynCalc symbolic software  package amounting to:
\begin{eqnarray}\label{m2nkr12g}
\overline{\vert{\mathcal M}\vert^2}&=&
\frac{1}{162 m^6 \left(m^2-s\right)^2 \left(m^2-u\right)^2}
\sum_{k=1}^7 m^{2k} a_{2k},
\end{eqnarray}
where $s,u$ are the standard Mandelstam variables and we are using the
notations
\begin{eqnarray}
a_0&=&18 s^2 u^2 (s+u)^3,\\
a_2&=&-9 s u (s+u)^2 (7 (s^2+u^2)+8 s u),\\
a_4&=&(s+u) (63 (s^4+u^4)\nonumber\\
&+&348 (s^3 u+s u^3)+578 s^2 u^2),\\
a_6&=&-165 (s^4+u^4)-588 (s^3 u+s u^3)-574 s^2 u^2,\\
a_8&=&2 (s+u) (5 (s^2+u^2)-142 s u),\\
a_{10}&=&2 (105 (s^2+u^2)-158 s u),\\
a_{12}&=&-280 (s+u),\\
a_{14}&=&912.
\end{eqnarray}
The differential cross section in the laboratory frame is now calculated from
the standard formulas, 
\begin{eqnarray}
\frac{d\s}{d\Omega}&=&\(\frac{1}{8\pi m}\frac{\o'}
{\o}\)^2\overline{\vert{\mathcal M}\vert^2},\\
\o'&=&\frac{m\o}{m+(1-\cos\q)\o},
\end{eqnarray}
where $\o$ and $\o'$ are the energies of the incident and scattered photons
respectively, while  $\q$ is
the scattering angle in the laboratory frame. Furthermore, with
\begin{eqnarray}
s&=&m(m+2\o),\\
u&=&m(m-2\o'),
\end{eqnarray}
and after some algebraic manipulations,  the final result is cast in the  form of an expansion in powers of $\eta^k$ (with $\eta=\omega/m$) according to,
\begin{equation}\label{dss}
\frac{d \s(\eta,x)}{d \Omega}=\frac{r_0^2}{162 (\eta  (x-1)-1)^5}
\sum_{k=0}^6 \eta ^k b_k.
\end{equation}
Here,  $r_0=e^2/(4\pi m)=\a/ m$, $x=\cos\q$, and the expansion coefficients are,
\begin{eqnarray}
b_0&=&-81 (x^2+1),\\
b_1&=&243 (x-1) (x^2+1),\\
b_2&=&-(x-1) (243 x^3-333 x^2+338 x-468),\\
b_3&=&(x-1)^2 (81 x^3-261 x^2+271 x-531),\\
b_4&=&(x-1)^2 (90 x^3-233 x^2+440 x-459),\\
b_5&=&6 (x-1)^3 (8 x^2-20 x+39),\\
b_6&=&9 (x-1)^3 (x^2-5 x+8).
\end{eqnarray}
In the low energy limit, we recover as expected the correct expression for the differential
cross section in the Thompson limit as,
\begin{equation}
\lim_{\eta \rightarrow 0}\frac{d \s(\eta ,x)}{d \Omega}=\frac{1}{2} r_0^2
\left(x^2+1\right),
\end{equation}
while in forward direction, the differential cross section takes an energy
independent value,
\begin{equation}
\lim_{x\rightarrow 1} \frac{d \s(\eta ,x)}{d \Omega}=r_0^2,
\label{unt_FD}
\end{equation}
and in accord with unitarity.
In all the other directions however, the differential cross section increases with 
energy. In the Figure \ref{dsfig} we present a plot of the quantity 
\begin{equation}
d\tilde{\s}=\frac{1}{r_0^2}\frac{d\s(\eta, x)}{d\Omega}
\end{equation}
as a function of the $x=\cos \theta$ variable, at energies of $\eta=0$ (solid curve), $\eta=1$ (long dashed curve) and $\eta=2.5$ (short dashed curve), here we see how the differential cross section approaches the classical limit at low energy (symmetric curve) and raises as the energy grows except in the forward direction.

Integrating over the solid angle we find the total cross section as:
\begin{equation}\label{tcs}
\s(\eta )=\sum_{k=0}^8 \frac{\eta ^k c_k\s_T}{108 \eta ^2 (2 \eta +1)^4}
+\sum_{\ell=0}^4 \frac{\eta ^\ell h_\ell\s_T \log (2 \eta +1)}{216 \eta ^3},
\end{equation}
being $\s_T=(8/3)\pi r_0^2$ the Thompson cross section and
\begin{eqnarray}
c_0&=&162, \quad c_1=1566,\\
c_2&=&6217,\quad c_3=12796,\\
c_4&=&14244,\quad c_5=8011,\\
c_6&=&1794,\quad c_7=126,\\
c_8&=&72,\quad h_0=-162,\\
h_1&=&-432,\quad h_2=-277,\\
h_3&=&-21,\quad h_4=27.
\end{eqnarray}
The total cross section (\ref{tcs})  has the following limits,
\begin{eqnarray}
\lim_{\eta \rightarrow 0}\s(\eta )&=&\s_T,\\
\lim_{\eta \rightarrow
\infty}\s(\eta )&=&\infty.
\end{eqnarray}

\begin{figure}
\includegraphics[scale=0.65]{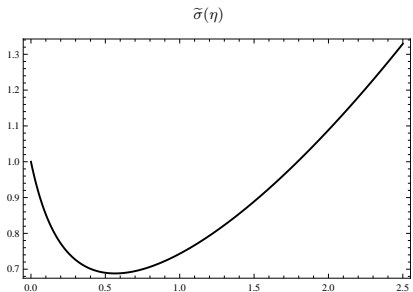}
\caption{\label{sfig} The total cross section $\widetilde{\sigma }(\eta )$ as a function
of $\eta =\omega /m$. In the low energy limit the Thompson limit,
$\widetilde{\sigma }(0)=1$, is recovered, otherwise the cross section
grows with the energy increase.}
\end{figure}

This behavior is show in Figure \ref{sfig}, where we make a plot of
\begin{equation}
\tilde{\s}=\frac{\s(\eta)}{\s_T},
\end{equation}
here we see the decreasing behavior of the cross section at low energies as well as its growing behavior at high energies.   

\section{Conclusions}
The first achievement of this  work is the representation in the equations~(\ref{DRCT})-(\ref{spin32ATS}) of the spin-$\frac{3}{2}$ degrees of freedom spanning the 
$\left(\frac{3}{2},0 \right)\oplus \left(0,\frac{3}{2} \right)$ carrier space of the Lo\-rentz algebra as anti-symmetric second rank 
Lorentz tensor-spinors. This representation is equipped by  separate Lo\-rentz and Dirac indexes and provides a 
comfortable  tool in calculations of scattering cross sections by means of the symbolic software package FeynCalc.
A similar experience has been made in  \cite{DelgadoAcosta:2013} regrading spin-$1$ transforming as 
$(1,0)\oplus(0,1)$, where the calculation of Compton scattering  could not be tackled in terms of six-dimensional spinors and was instead easily executed in the
anti-symmetric tensor-basis.
We here specifically worked out the Compton scattering off  $\left(\frac{3}{2},0 \right)\oplus \left(0,\frac{3}{2} \right)$ and reported in eq.~ (\ref{unt_FD}) 
on finite differential cross section in forward direction in the ultra relativistic regime and in accord with unitarity.
Before we had calculated in the eqs.~(\ref{QE0})--(\ref{QM3})  the multipole moments of a particle transforming according to the representation space
under investigation and found that they reproduced results earlier predicted by the Weinberg-Joos theory, the gyromagnetic ratio coming out fixed to the inverse of the spin, i.e. $g=\frac{1}{j}=\frac{2}{3}$ and in accord with Belinfante's conjecture. It is this very same value that gives rise to unitarity in the process of Compton scattering
in forward direction and at variance with the $g=2$ value required by spin-$\frac{3}{2}$ transforming in the four-vector spinor. 
Our observation suggests that spin-$\frac{3}{2}$ particles  transforming within the two-spin valued  four-vector spinor, on the one side, and
as the single spin valued $\left(\frac{3}{2},0 \right)\oplus \left(0,\frac{3}{2} \right)$, on the other side, behave as distinct species.
Within the framework of the second order approach elaborated here, we  
provided arguments in favor of the causal propagation of a pure spin-$\frac{3}{2}$ interacting  with an electromagnetic field.
We furthermore explained how our scheme allows for an extension toward any spin, be it bosonic or fermionic, in retaining the quadratic wave equations and 
associated Lagrangians.
Our conclusion is that $(j,0)\oplus (0,j)$ representation spaces  preserve their individuality upon  embeddings into 
reducible Lorentz-tensors. We have checked that also all the other irreducible sectors of the antisymmetric tensor-spinor, namely, 
the single spin-$\frac{1}{2}$ Dirac sector,  as well as the 
double spin valued $\left(1,\frac{1}{2} \right)\oplus \left(\frac{1}{2},1 \right)$ Rarita-Schwinger, 
are characterized before and after embeddings by same sets of multiple moments and Compton scattering cross sections.
 The  embeddings under discussion bring the advantage of separate Lorentz and  
 Dirac indexes which significantly speeds up the tensor calculus relatively to the matrix calculus.




\section*{Appendix: The explicit spin- $\frac{3}{2}$ degrees of freedom  contained within  the anti-symmetric tensor spinor}

As long as according to (\ref{tensor_spinor}) the anti-symmetric tensor spinor represents itself as the direct sum of 
$\left( \frac{3}{2},0\right)\oplus \left(0,\frac{3}{2} \right)$ ($8$ degrees of freedom)
with a Dirac spinor, $\left(\frac{1}{2},0\right)\oplus \left(0,\frac{1}{2}\right)$ (4 degrees of freedom), and
$\left(\frac{1}{2},1\right)\oplus \left(1,\frac{1}{2}\right)$ (12 degrees of freedom), the number of degrees of freedom is this space is $24$.
The set of the first  $16$  four-vector spinor degrees of freedom  has been  constructed in \cite{Napsuciale:2006wr}, 
with  those corresponding to spin-$\frac{3}{2}$ in 
$\left( \frac{1}{2},1\right) \oplus \left( 1,\frac{1}{2}\right)$ being

\begin{eqnarray} 
{\mathcal U}^\a_\pm\left(\mathbf{p},\frac{3}{2},\frac{3}{2}\right)&=&\eta^\a(\mathbf{p},1,1)u_\mp\left(\mathbf{p},\frac{1}{2}\right),\nonumber \\ \relax
{\mathcal U}_\pm^\a\left(\mathbf{p},\frac{3}{2},\frac{1}{2}\right)&=&\sqrt{\frac{1}{3}}\eta^\a(\mathbf{p},1,1)u_\mp\left(\mathbf{p},-\frac{1}{2}\right)\nonumber\\
&+&\sqrt{\frac{2}{3}}\eta^\a(\mathbf{p},1,0)u_\mp\left(\mathbf{p},\frac{1}{2}\right), \nonumber\\ \relax
{\mathcal U}^\a_\pm\left(\mathbf{p},\frac{3}{2},-\frac{1}{2}\right)&=&\sqrt{\frac{1}{3}}\eta^\a(\mathbf{p},1,-1)u_\mp\left(\mathbf{p},\frac{1}{2}\right)\nonumber\\
&+&\sqrt{\frac{2}{3}}\eta^\a(\mathbf{p},1,0)u_\mp\left(\mathbf{p},-\frac{1}{2}\right), \nonumber \\ \relax
{\mathcal U}_\pm^\a\left(\mathbf{p},\frac{3}{2},-\frac{3}{2}\right)&=&\eta^\a(\mathbf{p},1,-1)u_\mp\left(\mathbf{p},-\frac{1}{2}\right). \label{fvs324}
\end{eqnarray}
Here $u_\mp(\mathbf{p},\l)$ are negative and positive energy  Dirac spinors, coupled to  the following spin-$1$ four-vectors,
$\eta ^\alpha (\mathbf{p},\lambda)$ spanning $\left(\frac{1}{2},\frac{1}{2} \right)$, given (modulo notational differences) by \cite{Ahluwalia:2001}
\begin{eqnarray}
\eta^\a(\mathbf{p},1,1)&=&  \frac{1}{\sqrt{2}m(m+p_0)}\begin{pmatrix}
-(m+p_0)(p_1+ip_2) \\
-m^2-p_0m-p_1^2-ip_1p_2 \\
-i(p_2^2-ip_1p_2+m(m+p_0)) \\
-(p_1+ip_2)p_3
\end{pmatrix},\label{eta1pl} \\
\eta^\a(\mathbf{p},1,0)&=& \frac{1}{m(m+p_0)}
\begin{pmatrix}
(m+p_0)p_3 \\
p_1p_3 \\
p_2p_3 \\
p_3^2+m(m+p_0)
\end{pmatrix}, \label{eta10}\\
\eta^\a(\mathbf{p},1,-1)&=& \frac{1}{\sqrt{2}m(m+p_0)}\begin{pmatrix}
(m+p_0)(p_1-ip_2) \\
m^2+p_0m+p_1^2-ip_1p_2 \\
-i(p_2^2+ip_1p_2+m(m+p_0)).\\
(p_1-ip_2)p_3
\end{pmatrix}
\label{eta1mn} 
\end{eqnarray}
There is one more  four-vector residing within  the $\left( \frac{1}{2},\frac{1}{2}\right)$ representation space which is of 
spin-$0$ and reads \cite{Ahluwalia:2001},
\begin{equation}
\mbox{spin}-0 \in \left( \frac{1}{2},\frac{1}{2}\right):\quad \eta^\a(\mathbf{p},0,0)=\frac{p^\a}{m}.
\label{eta0}
\end{equation}
The coupling of the latter vector  to ${\mathcal U}^\a_\pm\left(\mathbf{p},\frac{3}{2},\lambda \right)$ in the equations (\ref{fvs324})  from above 
provides  a new independent  orthogonal complete set of spin-$\frac{3}{2}$  states according to, 
\begin{eqnarray}
\left[U_\pm\left(\mathbf{p},\frac{3}{2},\l\right)\right]^{\left[\a\b\right]}&=&
\frac{1}{2}{\Big(} \eta^\a(\mathbf{p},0,0){\mathcal U}_\pm^\b \left(\mathbf{p},\frac{3}{2},\l\right)
 - \eta^\b(\mathbf{p},0,0){\mathcal U}^\a _\pm\left(\mathbf{p},\frac{3}{2},\l\right) {\Big)}\nonumber\\
\label{ss32ATSprl}\\ 
&=&\frac{1}{2}{\Big( }\frac{p^\a}{m}{\mathcal U}^\b _\pm\left(\mathbf{p},\frac{3}{2},\l\right)
 - \frac{p^\b}{m}{\mathcal U}_\pm^\a \left(\mathbf{p},\frac{3}{2},\l\right){\Big)}.
\label{ss32ATS}
\end{eqnarray}
These are the tensor-spinors which, according to (\ref{spin32ATS}),  find themselves at the root of the degrees of freedom of the
$\left(\frac{3}{2},0\right)\oplus \left(0,\frac{3}{2} \right)$ representation space.
Finally, the conjugate  states to 
$ \overline{\mathcal U}_\pm^\a \left({\mathbf p},\frac{3}{2},\l\right) $, and
 $\left[ \overline{U}_\pm \left({\mathbf p},\frac{3}{2},\l\right)\right]^{\left[\a\b\right]}$ are defined as
\begin{eqnarray}
 \overline{{\mathcal U}}_\pm ^\a\left({\mathbf p},\frac{3}{2},\l\right)  &=& 
\left[ \gamma_0 {\mathcal U}_\pm \left({\mathbf p},\frac{3}{2},\l\right)\right]^{\dagger\a }, \label{dualdef1} \\ 
\left[ \overline{U}_\pm \left({\mathbf p},\frac{3}{2},\l\right)\right]^{\left[\a\b\right]} &=& 
\left[\gamma_0 U_\pm \left({\mathbf p},\frac{3}{2},\l\right)\right]^{\dagger {\left[\a\b\right]}}. \label{dualdef2}
\end{eqnarray}

\end{document}